\documentclass[a4paper]{IEEEtran}

\usepackage{etex}
\usepackage{graphicx,ifthen}
\usepackage{psfrag,amssymb,amsthm}
\usepackage[cmex10]{amsmath}

\usepackage{mathtools}
\usepackage[noadjust]{cite}

\usepackage{amsmath}
\usepackage{amsthm}
\usepackage{amssymb}
\usepackage{bm}
\usepackage{xspace}
\usepackage{xcolor}
\usepackage{url}
\usepackage{framed}
\usepackage{float}
\usepackage{rotating}
\usepackage{verbatim}
\usepackage{listings}
\usepackage{lscape}
\usepackage{pstricks,pst-sigsys,pst-plot}
\usepackage{subfigure}
\usepackage{algpseudocode,algorithm,algorithmicx}

 \usepackage{pgfplots}
 \usepackage{tikz}
 \usetikzlibrary{arrows,shapes.misc,chains,scopes,fit}

 \usepackage{booktabs}
 \usepackage{colortbl}

\newtheorem{lem}{Lemma}
\newtheorem{cor}{Corollary}
\newtheorem{prop}{Proposition}

\theoremstyle{definition}
\newtheorem{definition}{Definition}

\title{A Generalized Framework for Kullback-Leibler Markov Aggregation}

\author{
\IEEEauthorblockN{Rana Ali Amjad, Clemens Bl\"ochl, Bernhard C. Geiger}\\
\IEEEauthorblockA{Institute for Communications Engineering, Technical University of Munich, Germany\\
ranaali.amjad@tum.de, bernhard.geiger@tum.de}
}

\newcommand{\ent}[1]{H(#1)}

\newcommand{\mutinf}[1]{I(#1)}

\newcommand{\kld}[2]{D(#1||#2)}
\newcommand{\kldrate}[2]{\bar{D}(\mathbf{#1}||\mathbf{#2})}
\newcommand{\kldr}[2]{\bar{D}({#1}||{#2})}
\newcommand{\entrate}[1]{\bar{H}(\mathbf{#1})}

\newcommand{\redrate}[1]{\bar{R}(\mathbf{#1})}

\newcommand{\loss}[2][\empty]{\ifthenelse{\equal{#1}{\empty}}{L(#2)}{L_{#1}(#2)}}
\newcommand{\lossrate}[2][\empty]{\ifthenelse{\equal{#1}{\empty}}{L(\mathbf{#2})}{L_{\mathbf{#1}}(\mathbf{#2})}}

\newcommand{\relLoss}[2][\empty]{\ifthenelse{\equal{#1}{\empty}}{l(#2)}{l_{#1}(#2)}}

\DeclareMathOperator*{\argmin}{arg\,min}

\newcommand{\dom}[1]{\mathcal{#1}}

\newcommand{\Prob}[1]{\mathrm{Pr}(#1)}
\newcommand{\Mar}[1]{\mathrm{Mar}(#1)}

\newcommand{\e}[1]{\mathrm{e}^{#1}}

\newcommand{\card}[1]{\mathrm{card}(#1)}

\newcommand{\Xvec}{\mathbf{X}}

\newcommand{\Yvec}{\mathbf{Y}}

\newcommand{\Zvec}{\mathbf{Z}}

\newcommand{\Pvec}{\mathbb{P}}
\newcommand{\muvec}{\boldsymbol{\mu}}

\newcommand{\Wvec}{\mathbb{W}}

\newcommand{\Amat}{\mathbb{A}}

\newcommand{\onevec}{\mathbf{1}}

\newcommand{\eye}{\mathbf{I}}

\newcommand{\preim}[1]{g^{-1}(#1)}

\newcommand{\limn}{\lim_{n\to\infty}}

\newcommand{\Qvec}{\mathbb{Q}}
\newcommand{\Emat}{\mathbb{E}}
\newcommand{\nuvec}{\boldsymbol{\nu}}

\newcommand{\Uvec}{\mathbb{U}}
\newcommand{\Rvec}{\mathbb{R}}

\newcommand{\Marx}[1]{\tilde{#1}}
\renewcommand{\entrate}[1]{\bar{H}(#1)}
\renewcommand{\redrate}[1]{\bar{R}(#1)}
\renewcommand{\card}[1]{|#1|}

\newcommand{\pmf}[1]{p_{#1}}

\DeclareMathOperator{\diag}{diag}
\DeclareMathOperator{\diff}{\mathrm{d}}

\newcommand{\numitermax}{\text{\#iter}_{\max}}
\newcommand{\numiter}{\text{\#iter}}
\newcommand{\betatar}{\beta_{\text{target}}}

\newcommand{\epsrel}{\dom{R}_\varepsilon}

\newcommand{\ccost}[1]{\cost_{#1}(\Xvec,\Wvec)}
\newcommand{\costl}{\ccost{L}}
\newcommand{\costp}{\ccost{P}}
\newcommand{\costbeta}{\ccost{\beta}}
\newcommand{\ccostg}[1]{\cost_{#1}(\Xvec,g)}
\newcommand{\costlg}{\ccostg{L}}
\newcommand{\costpg}{\ccostg{P}}
\newcommand{\costbetag}{\ccostg{\beta}}

\begin{document}
\maketitle

\begin{abstract}
This paper proposes an information-theoretic cost function for aggregating a Markov chain via a (possibly stochastic) mapping. The cost function is motivated by two objectives: 1) The process obtained by observing the Markov chain through the mapping should be close to a Markov chain, and 2) the aggregated Markov chain should retain as much of the temporal dependence structure of the original Markov chain as possible. We discuss properties of this parameterized cost function and show that it contains the cost functions previously proposed by Deng et al., Xu et al., and Geiger et al.\ as special cases. We moreover discuss these special cases providing a better understanding and highlighting potential shortcomings: For example, the cost function proposed by Geiger et al.\ is tightly connected to approximate probabilistic bisimulation, but leads to trivial solutions if optimized without regularization. We furthermore propose a simple heuristic to optimize our cost function for deterministic aggregations and illustrate its performance on a set of synthetic examples.
\end{abstract}

\section{Introduction}\label{sec:intro}
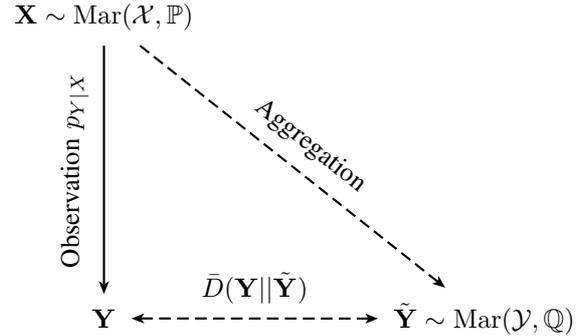
\begin{figure}[t] 
\centering
  \begin{pspicture}[showgrid=false](-1,1)(6,5.25)
    \psset{style=Arrow}
    \pssignal(0,5){x}{$\Xvec\sim\Mar{\dom{X},\Pvec}$}
    \pssignal(0,1){y}{$\Yvec$}
    \pssignal(5,1){yp}{$\Marx{\Yvec}\sim\Mar{\dom{Y},\Qvec}$}
    \ncline{<-}{y}{x} \aput{:U}{Observation $\pmf{Y|X}$}
    \ncline[style=Dash]{x}{yp}\aput{:U}{Aggregation}
    \pnode(4.8,1.4){yp1}\pnode(4.8,4.6){xp1}\pnode(5.2,1.4){yp2}\pnode(5.2,4.6){xp2}
    \ncline[style=Dash]{<->}{y}{yp}\Aput{$\kldr{\Yvec}{\Marx{\Yvec}}$}
\end{pspicture}
\caption{Illustration of the aggregation problem: A stationary first-order Markov chain $\Xvec$ is given. We are interested in finding a conditional distribution $\pmf{Y|X}$ and an aggregation of $\Xvec$, i.e., a  Markov chain $\Marx{\Yvec}$ on $\dom{Y}$. The conditional distribution $\pmf{Y|X}$ defines a stationary process $\Yvec$, a noisy observation of $\Xvec$. $\Yvec$ might not be Markov of any order, but can be approximated by a Markov chain $\Marx{\Yvec}$.}
\label{fig:aggregation}
\end{figure}

Markov aggregation is the task of representing a Markov chain with a large alphabet by a Markov chain with a smaller alphabet, thus reducing model complexity while at the same time retaining the computationally and analytically desirable Markov property (see Fig.~\ref{fig:aggregation}). Such a model reduction is necessary if the original Markov chain is too large to admit simulation, estimating model parameters from data, or control (in the case of Markov decision processes). These situations occur often in computational chemistry (where aggregation is called coarse-graining, e.g.,~\cite{Katsoulakis_CoarseGraining}), natural language processing, and the simulation and control of large systems (giving rise to the notion of bisimulation, e.g.,~\cite{Abate_Bisimulation}). Additionally, Markov aggregation can be used as a tool in exploratory data analysis, either to discover groups of ``similar'' states of a stochastic process or to cluster data points, cf.~\cite{Alush_PairwiseClustering,Tishby_MarkovRelaxation}.

Information-theoretic cost functions were proposed for Markov aggregation in~\cite{Meyn_MarkovAggregation,Xu_Reduction,GeigerEtAl_OptimalMarkovAggregation,Vidyasagar_MarkovAggregation}. Specifically, the authors of~\cite{Meyn_MarkovAggregation} proposed a cost function linked to the \emph{predictability} of the aggregated Markov chain. Such an approach is justified if the original model is \emph{nearly completely decomposable}, i.e., if there is a partition of the alphabet such that transitions within each element of the partition occur quickly and randomly, while transitions between elements of the partition occur only rarely. Building on this work, the authors of~\cite{GeigerEtAl_OptimalMarkovAggregation} proposed a cost function linked to \emph{lumpability}, i.e., to the phenomenon where a function of a Markov chain is Markov. Such an approach is justified whenever there are groups of states with similar probabilistic properties (in a well-defined sense). Both~\cite{Meyn_MarkovAggregation} and~\cite{GeigerEtAl_OptimalMarkovAggregation} focus on deterministic aggregations, i.e., every state of the original alphabet is mapped to exactly one state of the reduced alphabet. Moreover, the authors of both references arrive at their cost functions by lifting the aggregated Markov chain to the original alphabet. The authors of~\cite{Xu_Reduction} present an information-theoretic cost function for stochastic aggregations, but they do not justify their choice by an operational characterization (such as predictability or lumpability). Instead, they arrive at their cost function via the composite of the original and the aggregated Markov chain.

In this paper, we extend the works~\cite{Meyn_MarkovAggregation,Xu_Reduction,GeigerEtAl_OptimalMarkovAggregation,Vidyasagar_MarkovAggregation} as follows:
\begin{enumerate}
 \item We present a two-step approach to Markov aggregation (Section~\ref{sec:Aggregation}): Observing the original Markov chain through a (stochastic or deterministic) channel, and then approximating this (typically non-Markov) process as a Markov chain (see Fig.~\ref{fig:aggregation}). This approach has already been taken by~\cite{GeigerEtAl_OptimalMarkovAggregation}, albeit only for deterministic aggregations.
 \item Using this two-step approach, we propose a parameterized, information-theoretic cost function for Markov aggregation (Section~\ref{sec:regularized}). We arrive at this cost function neither via lifting nor via the composite model, but via requiring specific operational qualities of the process observed through the channel: It should be close to a Markov chain and it should retain the temporal dependence structure of the original Markov chain.
 \item We show that our cost function contains the cost functions of~\cite{Meyn_MarkovAggregation,Xu_Reduction,GeigerEtAl_OptimalMarkovAggregation,Vidyasagar_MarkovAggregation} as special cases (Section~\ref{sec:related}). We also discuss previous algorithmic approaches to the Markov aggregation problem.
 \item We propose a simple, low-complexity heuristic to minimize our generalized cost function for deterministic aggregations (Section~\ref{sec:algos}).
 \item As a side result, we justify the cost function proposed in~\cite{GeigerEtAl_OptimalMarkovAggregation} by showing a tight connection to \emph{approximate probabilistic bisimulation} (Section~\ref{ssec:lumpability}).
\end{enumerate}

We illustrate our cost function for various examples in Section~\ref{sec:experiments}. Specifically, we investigate the aggregation of quasi-lumpable and nearly completely decomposable Markov chains, and we look at a toy example from natural language processing. We also take up the approach of~\cite{Alush_PairwiseClustering,Tishby_MarkovRelaxation} to perform clustering via Markov aggregation. In future work, we shall extend our efforts to Markov decision processes, and provide a theory for lifting stochastic aggregations as indicated in~\cite[Remark~3]{Xu_Reduction}.

\section{Notation and Definitions}\label{sec:prelim}
We denote vectors and matrices by bold lower case and blackboard bold upper case letters, e.g., $\mathbf{a}$ and $\Amat$. A diagonal matrix with vector $\mathbf{a}$ on the main diagonal is denoted by $\diag(\mathbf{a})$. The transpose of $\Amat$ is $\Amat^T$.

We denote random variables (RVs) by upper case letters, e.g., $Z$, and their alphabet by calligraphic letters, e.g., $\dom{Z}$. In this work we will restrict ourselves to RVs with finite alphabets, i.e., $\card{\dom{Z}}<\infty$. Realizations are denoted by lower case letters, e.g., $z$, where $z\in\dom{Z}$. The probability mass function (PMF) of $Z$ is $p_Z$, where $p_Z(z) := \Prob{Z=z}$ for all $z\in\dom{Z}$. Joint and conditional PMFs are defined accordingly.

We denote a one-sided, discrete-time, stochastic process with  $\Zvec:=(Z_1,Z_2,\dots)$, where each $Z_k$ takes values from the (finite) alphabet $\dom{Z}$. We abbreviate $Z_m^n:=(Z_m,Z_{m+1},Z_n)$. We consider only \emph{stationary} processes, i.e., PMFs are invariant w.r.t.\ a time shift. In particular, the marginal distribution of $Z_k$ is equal for all $k$ and shall be denoted as $p_Z$.

A first-order Markov chain is a process that satisfies, for all $n>1$ and all $z_1^n\in\dom{Z}^n$,
\begin{align}\label{eq:Markovdef}
 \pmf{Z_n|Z_1^{n-1}}(z_n|z_1^{n-1}) &=  \pmf{Z_n|Z_{n-1}}(z_n|z_{n-1}).
\end{align}
The Markov chain is time-homogeneous if, the right-hand side of~\eqref{eq:Markovdef} does not depend on $n$, i.e., if
\begin{align}
\pmf{Z_n|Z_{n-1}}(z_n|z_{n-1}) &= \pmf{Z_2|Z_{1}}(z_n|z_{n-1}):=P_{z_{n-1}\to z_n}.
\end{align}
If the transition probability matrix $\Pvec=[P_{z_{n-1}\to z_n}]$ of a time-homogeneous Markov chain is irreducible and aperiodic (see~\cite{Kemeny_FMC} for terminology), then there exists a unique vector $\muvec$ such that $\muvec^T=\muvec^T\Pvec$, which represents the invariant distribution of $\Zvec$. If the initial distribution $\pmf{Z_1}$ coincides with $\muvec$, then $\Zvec$ is stationary and we denote this stationary, irreducible and aperiodic first-order Markov chain by $\Zvec\sim\Mar{\dom{Z},\Pvec}$. In this work we deal exclusively with first-order stationary, irreducible and aperiodic Markov chains.

We use information-theoretic cost functions for the aggregation problem. The entropy of $Z$, the conditional entropy of $Z_2$ given $Z_1$, and the mutual information between $Z_1$ and $Z_2$ are defined by
\begin{subequations}
 \begin{align}
  \ent{Z} &:= -\sum\limits_{z\in\dom{Z}} \pmf{Z}(z) \log \pmf{Z}(z)\\
  \ent{Z_2|Z_1} &:= \sum\limits_{z\in\dom{Z}} \ent{Z_2|Z_1=z} \pmf{Z_1}(z)\\
  \mutinf{Z_1;Z_2} &:= \ent{Z_2} - \ent{Z_2|Z_1}.
 \end{align}
The entropy rate and redundancy rate of a stationary stochastic process $\Zvec$ (not necessarily Markov) are
\begin{align}\label{eq:ERdef}
 \entrate{\Zvec} &:= \limn \frac{\ent{Z_1^{n}}}{n} = \limn \ent{Z_n|Z_{1}^{n-1}}\\
 \redrate{\Zvec} &:= \limn \mutinf{Z_n;Z_1^{n-1}} = \ent{Z}-\entrate{\Zvec}.
\end{align}
The Kullback-Leibler divergence rate (KLDR) between two stationary stochastic processes $\Zvec$ and $\Zvec'$ on the same finite alphabet $\dom{Z}$ is~\cite[Ch.~10]{Gray_Entropy}
\begin{equation}\label{eq:KLDRdef}
 \kldrate{Z'}{Z} := \limn \frac{1}{n}\sum\limits_{z_1^n\in\dom{Z}^n} p_{{Z'}_1^n}(z_1^n) \log \frac{p_{{Z'}_1^n}(z_1^n)}{p_{Z_1^n}(z_1^n)}
\end{equation}
provided the limit exists. If the limit exists, it is finite if, for all $n$ and all $z_1^n$, $p_{Z_1^n}(z_1^n) = 0$ implies $p_{{Z'}_1^n}(z_1^n) = 0$ (short: $p_{{Z'}_1^n}\ll p_{Z_1^n}$). In particular, if $\Zvec'\sim\Mar{\dom{Z},\Pvec'}$ and $\Zvec\sim\Mar{\dom{Z},\Pvec}$, then~\cite{Rached_KLDR}
\begin{equation}\label{eq:KLDRMarkov}
 \kldrate{Z'}{Z}= \sum\limits_{z,z'\in\dom{Z}} \mu_z P'_{z\to z'}\log\frac{P'_{z\to z'}}{P_{z\to z'}}.
\end{equation}
provided $\Pvec\ll\Pvec'$.
\end{subequations}

These information-theoretic quantities can be used to give an equivalent definition of Markovity:
\begin{lem}[{\cite[Prop.~3]{GeigerTemmel_kLump}}]\label{lem:ITMarkov}
Suppose the stochastic process $\Zvec$ is stationary. Then, $\Zvec$ is Markov iff $\entrate{\Zvec}=\ent{Z_2|Z_1}$.
\end{lem}

If $\Zvec$ is a stationary process on $\dom{Z}$ (not necessarily Markov), then one can approximate this process by a Markov chain $\Marx{\Zvec}\sim\Mar{\dom{Z},\Pvec}$:

\begin{lem}[{\cite[Cor.~10.4]{Gray_Entropy}}]\label{lem:MarkovApprox}
	Let $\Zvec$ be a stationary process on $\dom{Z}$, and let $\Zvec'\sim\Mar{\dom{Z},\Pvec'}$ be any Markov chain on $\dom{Z}$. Then,
	\begin{subequations}\label{eq:Approx}
		\begin{equation}
		\Pvec=\argmin_{\Pvec'} \kldr{\Zvec}{\Zvec'}
		\end{equation}
		where
		\begin{equation}\label{eq:MApprox}
		P_{z\to z'} = \pmf{Z_2|Z_1}(z'|z).
		\end{equation}
		Moreover, for $\Marx{\Zvec}\sim\Mar{\dom{Z},\Pvec}$,
		\begin{equation}\label{eq:ApproxError}
		\kldrate{\Zvec}{\Marx{\Zvec}} = \ent{Z_2|Z_1}-\entrate{\Zvec}.
		\end{equation}
	\end{subequations}
\end{lem}

By Lemma~\ref{lem:ITMarkov} we know that right-hand side of~\eqref{eq:ApproxError} is 0 iff $\Zvec$ is Markov. Hence, one can view the KLDR $\kldrate{\Zvec}{\Marx{\Zvec}}$ as a measure of how close a process $\Zvec$ is to a Markov chain.

\section{Markov Chain Aggregation}\label{sec:Aggregation}
\newcommand{\cost}{\dom{C}}
Given a Markov chain $\Xvec$, Markov aggregation deals with the problem of finding a Markov chain $\Marx{\Yvec}$ on a given smaller alphabet $\dom{Y}$ which is the optimal representation of $\Xvec$ in the sense of minimizing a given cost function $\bar{\cost}(\Xvec,\Marx{\Yvec})$. This is depicted in Fig.~\ref{fig:aggregation} by the diagonal arrow and is summarized in the following definition:
\begin{definition}[Markov Aggregation Problem]\label{def:problem}
	Let $\Xvec\sim\Mar{\dom{X},\Pvec}$, $\dom{Y}$, and an arbitrary cost function $\bar{\cost}(\cdot,\cdot)$ be given. The Markov aggregation problem concerns finding a minimizer of
	\begin{equation}\label{eq:genericcost}
	\min_{\Marx{\Yvec}}\  \bar{\cost}(\Xvec,\Marx{\Yvec})
	\end{equation}
	where the optimization is over Markov chains on $\dom{Y}$.
\end{definition}
In this work we address the Markov aggregation problem using the two-step approach depicted in Fig~\ref{fig:aggregation}. The first step is to use a (possibly stochastic) mapping from $\dom{X}$ to $\dom{Y}$. Applying this mapping to $\Xvec$ leads to a stationary process $\Yvec$ which may not be Markov (in fact, $\Yvec$ is a hidden Markov process). In the second step we look for the optimal approximation $\Marx{\Yvec}$ of $\Yvec$ in the sense of Lemma~\ref{lem:MarkovApprox}.

This two-step approach is a popular method of Markov aggregation and has been employed in various works including \cite{Meyn_MarkovAggregation,GeigerEtAl_OptimalMarkovAggregation,Vidyasagar_MarkovAggregation}. In these references, the mapping in the first step was restricted to be deterministic whereas in this work we allow it to be stochastic. In other words, while these references were looking for a partition of $\dom{X}$ induced by a function $g{:}\ \dom{X}\to\dom{Y}$, in this work we permit stochastic mappings induced by a conditional distribution $\pmf{Y|X}$. We represent $\pmf{Y|X}$ as a row stochastic matrix $\Wvec=[W_{x\to y}]$, where $W_{x\to y}=\pmf{Y|X}(y|x)$. 

With this notation, the following corollary to Lemma~\ref{lem:MarkovApprox} solves the second of the two steps in our approach, i.e., it characterizes the optimal approximation $\Marx{\Yvec}$ of the hidden Markov process $\Yvec$ that we obtain by observing $\Xvec$ through $\Wvec$:
\begin{cor}\label{cor:MAgg}
	Let $\Xvec\sim\Mar{\dom{X},\Pvec}$ and let $\Wvec$ denote a conditional distribution from $\dom{X}$ to $\dom{Y}$. Let $\Yvec$ be the hidden Markov process obtained by observing $\Xvec$ through $\Wvec$, and let $\Marx{\Yvec}\sim\Mar{\dom{Y},\Qvec}$ be its best Markov approximation in the sense of minimizing $\kldr{\Yvec}{\Marx{\Yvec}}$ (cf.\ Lemma~\ref{lem:MarkovApprox}). Then,
	\begin{equation}\label{eq:Qagg}
	\Qvec = \Uvec\Pvec\Wvec
	\end{equation}
	where $\Uvec:=\diag(\nuvec)^{-1}\Wvec^T\diag(\muvec)$ with $\nuvec^T:=\muvec^T\Wvec$ being the marginal distribution of $Y_k$.
\end{cor}
Note that this corollary extends~\cite[Lem.~3]{GeigerEtAl_OptimalMarkovAggregation} from deterministic to stochastic mappings.

With the second step solved, the two-step approach to the optimization problem stated in Definition~\ref{def:problem} boils down to optimization over the mapping $\Wvec$. We can thus restate the Markov aggregation problem as follows:
\begin{definition}[Markov Aggregation Problem Restated]\label{def:problemrestated}
	Let $\Xvec\sim\Mar{\dom{X},\Pvec}$, $\dom{Y}$, and an arbitrary cost function $\bar{\cost}(\cdot,\cdot)$ be given. Let 
	\begin{equation}
	\cost(\Xvec,\Wvec) =  \bar{\cost}(\Xvec,\Marx{\Yvec})
	\end{equation}
	where $\Marx{\Yvec}$ is the Markov approximation of the hidden Markov process $\Yvec$ that is obtained by observing $\Xvec$ through the stochastic mapping $\Wvec$. The Markov aggregation problem using the two-step approach concerns finding a minimizer of
	\begin{equation}\label{eq:Wcost}
	\min_{\Wvec}\ \cost(\Xvec,\Wvec)
	\end{equation}
	where the optimization is over stochastic mappings from $\dom{X}$ to $\dom{Y}$. If the optimization is restricted over deterministic mappings $g$, we abuse notation and write $\cost(\Xvec,g)$ for the cost.
\end{definition}
 Note that Definition \ref{def:problem} and Definition \ref{def:problemrestated} are not equivalent in general, i.e., the optimal aggregated chain $\Marx{\Yvec}$ obtained by solving \eqref{eq:genericcost} is not the same as the optimal aggregated chain $\Marx{\Yvec}$ obtained by solving \eqref{eq:Wcost}. The two formulations only become equivalent when we restrict the optimization in \eqref{eq:genericcost} to aggregated Markov chains which can be obtained as a result of the aforementioned two-step approach.
 
\subsection{Markov Aggregation via Lumpability}\label{ssec:lumpability}
The optimal mapping $\Wvec$ depends on the cost function $\bar{\cost}$. One possible choice in the light of Lemma~\ref{lem:MarkovApprox} is
\begin{equation}\label{eq:KLDRcost}
 \bar{\cost}(\Xvec,\Marx{\Yvec})=\kldr{\Yvec}{\Marx{\Yvec}}.
\end{equation}
In other words, we wish to find a mapping $\Wvec$ such that the hidden Markov process $\Yvec$ is as close to a Markov chain as possible in an information-theoretic sense. This may be reasonable since it states that data obtained by simulating the aggregated model $\Marx{\Yvec}$ differs not too much from data obtained by simulating the original model in conjunction with the stochastic mapping, i.e., data obtained from $\Yvec$.

There are two shortcomings of the cost~\eqref{eq:KLDRcost}. First,~\eqref{eq:KLDRcost} focuses only on getting $\Yvec$ close to $\tilde{\Yvec}$ but not on preserving any form of information in $\Xvec$. This gives rise to trivial optimal solutions: If $\Wvec$ is such that the conditional distribution  does not depend on the conditioning event, i.e., $\pmf{Y|X}=\pmf{Y}$, or $\Wvec=\onevec\boldsymbol{\alpha}^T$ for some probability vector $\boldsymbol{\alpha}$, then $\Yvec$ is independent and identically distributed (i.i.d.) and hence Markov. Indeed, in this case $\entrate{\Yvec}=\ent{Y_2|Y_1}=\ent{Y}$, from which $\kldr{\Yvec}{\Marx{\Yvec}}=0$ follows. The cost function is thus inappropriate for Markov aggregation, unless it is regularized appropriately. The second shortcoming is linked to the fact that $\kldr{\Yvec}{\Marx{\Yvec}}$ requires, by~\eqref{eq:ApproxError}, the computation of the entropy rate $\entrate{\Yvec}$ of a hidden Markov process. This problem is inherently difficult~\cite{Blackwell_HMMRate}, and analytic expressions do not even exist for simple cases (cf.~\cite{Ordentlich_LowerBound}). In the following, we discuss two previously proposed relaxations of the Markov aggregation problem for $\bar{\cost}(\Xvec,\Marx{\Yvec})=\kldr{\Yvec}{\Marx{\Yvec}}$.

The authors of~\cite{GeigerEtAl_OptimalMarkovAggregation} addressed the second shortcoming by relaxing the cost via
\begin{equation}\label{eq:lumpCost}
 \costl:=\ent{Y_2|Y_1} - \ent{Y_2|X_1}\ge \kldr{\Yvec}{\Marx{\Yvec}}.
\end{equation}
This cost does not require computing $\entrate{\Yvec}$ and is linked to the phenomenon of \emph{lumpability}, the fact that a function of a Markov chain has the Markov property~\cite[Thm.~9]{GeigerTemmel_kLump}: If $\costl=0$, then $\Yvec$ is a Markov chain.

We now show that, at least for deterministic mappings, this cost function also has a justification in \emph{approximate probabilistic bisimulations}, or $\varepsilon$-bisimulations. More specifically, the authors of~\cite{Desharnais_Bisimulation} discussed bisimilarity of Markov processes and showed that two Markov chains are bisimilar if one can be described as a function of the other (see discussion after~\cite[Def.~5.2]{Desharnais_Bisimulation}). In other words, if $\Xvec\sim\Mar{\dom{X},\Pvec}$ is a Markov chain, $g{:}\ \dom{X}\to\dom{Y}$ a surjective function, and $\Yvec\sim\Mar{\dom{Y},\Qvec}$ satisfies $Y_k=g(X_k)$, then $\Xvec$ and $\Yvec$ are bisimilar. Since this is equivalent to lumpability, bisimilarity is implied by $\costlg=0$.

Extending this line of reasoning, we give a justification of the cost function $\costlg$ in terms of $\varepsilon$-bisimulation of pairs of Markov chains, even in case $\Xvec$ is not lumpable w.r.t.\ $g$. To this end, we adapt~\cite[Def.~4~\&~5]{Bian_Bisimulation} for our purposes:
\begin{definition}[$\varepsilon$-Bisimulation]\label{def:bisimulation}
 Consider two finite Markov chains $\Xvec\sim\Mar{\dom{X},\Pvec}$ and $\Marx{\Yvec}\sim\Mar{\dom{Y},\Qvec}$ and assume w.l.o.g.\ that $\dom{X}$ and $\dom{Y}$ are disjoint. We say that $\Xvec$ and $\Marx{\Yvec}$ are \emph{$\varepsilon$-bisimilar} if there exists a relation $\epsrel\subseteq (\dom{X}\cup\dom{Y})\times(\dom{X}\cup\dom{Y})$ such that for all $x\in\dom{X}$ and $y\in\dom{Y}$ for which $(y,x)\in \epsrel$, and all $T\subseteq \dom{X}\cup\dom{Y}$ we have
 \begin{equation}
  \sum_{x'\in \epsrel(T)\cap\dom{X}}P_{x\to x'} \ge \sum_{y'\in T\cap\dom{Y}} Q_{y\to y'} - \varepsilon
 \end{equation}
  where $\epsrel(T):=\{s_2\in\dom{X}\cup\dom{Y}{:}\ s_1\in T, (s_1,s_2)\in \epsrel\}$.
\end{definition}

The definitions of $\varepsilon$-bisimulations are typically given for labeled~\cite[Def.~4~\&~5]{Bian_Bisimulation} or controlled~\cite[Def.~4.4]{Abate_Bisimulation} Markov processes with general alphabets and thus contain more restrictive conditions than our Definition~\ref{def:bisimulation}. Our definition is equivalent if the alphabets are finite and if the set of labels is empty. We are now ready to state
\begin{prop}\label{prop:bisimulation}
 Let $\Xvec\sim\Mar{\dom{X},\Pvec}$ and the surjective function $g{:}\ \dom{X}\to\dom{Y}$ be given. Let $\Qvec$ be as in Corollary~\ref{cor:MAgg}, where $W_{x\to y}=1$ iff $y=g(x)$. Let $\Marx{\Yvec}\sim\Mar{\dom{Y},\Qvec}$. Then, $\Xvec$ and $\Marx{\Yvec}$ are $\varepsilon$-bisimilar with
 \begin{equation}
  \varepsilon = \sqrt{\frac{\ln(2) \costlg}{2 \min_{x\in\dom{X}} \mu_x }}.
 \end{equation}
\end{prop}

 \begin{IEEEproof}
  See Section~\ref{proof:bisimulation}.
 \end{IEEEproof}

 Despite this justification, the cost function $\costl$ is mainly of theoretical interest. The reason is that the shortcoming of leading to trivial solutions is inherited by $\costl$, since for $\Wvec=\onevec\boldsymbol{\alpha}^T$ one gets $\costl=0$, regardless of $\boldsymbol{\alpha}$ and $\Pvec$. Even restricting $\Wvec$ to be a deterministic partition, as considered in \cite{GeigerEtAl_OptimalMarkovAggregation}, does not solve this problem: The combinatorial search over all partitions may have its global optimum at a partition that makes $\Yvec$ close to an i.i.d.\ process. Indeed, if the cardinality of $\dom{Y}$ is not constrained (or if $g$ is not required to be surjective), then the constant function $g$ yields $\costlg=0$.
 
 \subsection{Markov Aggregation by Predictability}\label{ssec:predictability}
A different approach was taken by the authors of~\cite{Meyn_MarkovAggregation} who proposed the following cost function (again with the focus on deterministic partitions):
\begin{equation}\label{eq:predCost}
 \costp:=\mutinf{X_1;X_2} - \mutinf{Y_1;Y_2}
\end{equation}
The computation of $\costp$ is simple as it does not require computing $\entrate{\Yvec}$. Furthermore, $\costp$ reflects the wish to preserve the \emph{temporal dependence structure} of $\Xvec$, i.e., it is connected to predicting future states of $\Yvec$ based on knowledge of past states of $\Yvec$. Since $\Xvec$ is not i.i.d., observing $X_k$ reveals some information about $X_{k+1}$. Minimizing $\costp$ thus tries to find a $\Wvec$ such that $Y_k$ reveals as much information about $Y_{k+1}$ as possible, and hence does not lead to the same trivial solutions as $\costl$ and $\kldr{\Yvec}{\Marx{\Yvec}}$:  A constant function $g$ or a soft partition $\Wvec=\onevec\boldsymbol{\alpha}^T$ render $Y_1$ and $Y_2$ independent, hence the cost is maximized at $\costp=\mutinf{X_1;X_2}$. Unfortunately, as it was shown in~\cite[Thm.~1]{GeigerEtAl_OptimalMarkovAggregation}, we have $\costp\ge\costl$, i.e.,~\eqref{eq:predCost} does not capture Markovity of $\Yvec$ as well as the relaxation proposed by~\cite{GeigerEtAl_OptimalMarkovAggregation}.

Note that although $\costp$ does not lead to the same trivial solutions as $\costl$ and $\kldr{\Yvec}{\Marx{\Yvec}}$, it still tries to preserve only part of the information about temporal dependence in $\Xvec$, i.e., information which is helpful in predicting the next sample. Such a goal is justified in scenarios in which $\Xvec$ is quasi-static, i.e., runs on different time scales: The process $\Xvec$ moves quickly and randomly within a group of states, but moves only slowly from one group of states to another. 

Since all other information contained in $\Xvec$ is not necessarily preserved by minimizing $\costp$, this cost function can also lead to undesired solutions: For example, if $\Xvec$ is i.i.d.\ and hence does not contain any temporal dependence structure, then $\mutinf{X_1;X_2}=0$ and $\costp=0$ for every mapping $\Wvec$. 

A third objective for optimization may be worth mentioning. The information contained in $\Xvec$ splits into a part describing its temporal dependence structure (measured by its redundancy rate $\mutinf{X_1;X_2}$) and a part describing its new information generated in each time step (measured by its entropy rate $\entrate{\Xvec}$). Indeed, we have
\begin{equation}\label{eq:allinfo}
  \ent{X} = \entrate{\Xvec}+  \mutinf{X_1;X_2}.
\end{equation}
While in this work we focus on preserving Markovity via $\costl$ and the temporal dependence structure  via $\costp$, the authors of~\cite{GeigerTemmel_kLump} investigated conditions such that the newly generated information (measured by $\entrate{\Xvec}$) is preserved. Developing a Markov aggregation framework that trades between three different goals -- Markovity, temporal dependence, generated information -- is the object of future work.

\section{Regularized Markov Chain Aggregation}\label{sec:regularized}
In this section we combine the approaches in \cite{GeigerEtAl_OptimalMarkovAggregation} and \cite{Meyn_MarkovAggregation} to obtain a new cost function for Markov aggregation. As discussed in Section~\ref{sec:Aggregation}, the aim of Markov aggregation in \cite{GeigerEtAl_OptimalMarkovAggregation} is to get a process $\Yvec$ which is as Markov as possible. This can be captured well by the cost function $\kldr{\Yvec}{\Marx{\Yvec}}$ in the light of Lemma \ref{lem:MarkovApprox}. The authors of \cite{Meyn_MarkovAggregation} define the Markov aggregation problem in terms of finding a mapping $\Wvec$ which preserves the temporal dependence structure in $\Xvec$. This temporal dependence is captured well by the redundancy rate of the process, which for a Markov chain $\Xvec$ equals $\redrate{\Xvec}=\mutinf{X_1;X_2}$. Preserving this temporal dependence structure is thus well captured by maximizing the redundancy rate of $\Yvec$, i.e., by the following optimization problem:
\begin{equation}
 \min_{\Wvec}\  \redrate{\Xvec}-\redrate{\Yvec}
\end{equation}
Hence, to combine both the goal of Markovity and the goal of preserving temporal information, one can define the following Markov aggregation problem
\begin{equation}\label{eq:gencost}
 \min_{\Wvec}\ \underbrace{(1-\beta)\kldr{\Yvec}{\Marx{\Yvec}} + \beta(\redrate{\Xvec}-\redrate{\Yvec})}_{=:\delta_\beta(\Xvec,\Wvec)}
\end{equation}
where $0 \leq \beta \leq 1$. Clearly, for $\beta=0$ we are back at Definition~\ref{def:problem}. For a general $\beta$, the data processing inequality ensures that $\redrate{\Xvec}\ge\redrate{\Yvec}$, hence the cost~\eqref{eq:gencost} is non-negative. We moreover have
\begin{lem}\label{lem:deltabeta_nondecreasing}
	$\delta_\beta(\Xvec,\Wvec)$ is non-decreasing in $\beta$.
\end{lem}
\begin{IEEEproof}
	See Section~\ref{proof:deltabeta_nondecreasing}.
\end{IEEEproof}
Although minimizing $\delta_\beta(\Xvec,\Wvec)$ tries to preserve both Markovity and the temporal information in $\Xvec$, the computation of \eqref{eq:gencost} requires computing $\entrate{\Yvec}$. We thus take the approach of \cite{GeigerEtAl_OptimalMarkovAggregation} to relax \eqref{eq:gencost}. By  ~\cite[Thm.~1]{GeigerEtAl_OptimalMarkovAggregation}, we have
\begin{equation}
\kldr{\Yvec}{\Marx{\Yvec}} \leq \costl.
\end{equation}
By the data processing inequality we also have 
\begin{equation}
\redrate{\Xvec}-\redrate{\Yvec} \leq \costp.
\end{equation}
Hence, combining the two we get an upper bound on $\delta_\beta(\Xvec,\Wvec)$ that does not require computing $\entrate{\Yvec}$. Indeed, for $\beta\in[0,1]$,
\begin{equation}\label{eq:costbound}
\delta_\beta(\Xvec,\Wvec)\le(1-\beta) \costl + \beta\costp.
\end{equation}
Rather than the right-hand side of~\eqref{eq:costbound}, we propose the following cost function for Markov aggregation:
\begin{equation}\label{eq:finalcost}
\costbeta := (1-2\beta) \costl + \beta\costp.
\end{equation}
where again $\beta\in[0,1]$. One can justify going from \eqref{eq:costbound} to \eqref{eq:finalcost} by noticing that for every $0 \leq \beta \leq 1$ for \eqref{eq:costbound}, one can find a $0 \leq \beta \leq 0.5$ for \eqref{eq:finalcost} such that the two optimization problems are equivalent, i.e., they have the same optimizer $\Wvec$. Furthermore, for $\beta =1$, the cost function in \eqref{eq:finalcost} corresponds to information bottleneck problem, a case that is not covered by \eqref{eq:costbound}. Hence, not only is $\cost_\beta$ a strict generalization of $\delta_\beta$ but also has the information bottleneck problem as an interesting corner case. In the following we summarize some of the properties of $\cost_\beta$.

\begin{lem}\label{lem:costbeta_properties}
For $\cost_\beta$ and $0 \leq \beta \leq 1$ we have:
	\begin{enumerate}
		\item $\costbeta\ge 0$
		\item $\delta_{0.5}(\Xvec,\Wvec)=\ccost{0.5}=\frac{1}{2}\costp$
		\item $\ccost{1}=\ccost{\mathrm{IB}}:=\mutinf{X_1;X_2|Y_2}$
		\item For $\beta\le \frac{1}{2}$, $\beta\costp\le\delta_\beta(\Xvec,\Wvec)\le\costbeta$
		\item For $\beta\ge\frac{1}{2}$, $\costbeta\le\delta_\beta(\Xvec,\Wvec)\le\beta\costp$
		\item If $\Xvec$ is reversible, then $\costbeta$ is non-decreasing in $\beta$
	\end{enumerate}
\end{lem}

\begin{IEEEproof}
	See Section~\ref{proof:costbeta_properties}.
\end{IEEEproof}

\section{Related Work: Special Cases of $\costbeta$}\label{sec:related}
We now show that specific settings of $\beta$ lead to cost functions that have been proposed previously in the literature. We list these approached together with the algorithms that were proposed to solve the respective Markov aggregation problem.
\begin{itemize}
	\item For $\beta=\frac{1}{2}$, optimizing \eqref{eq:finalcost} is equivalent to optimizing $\costp$. The authors of~\cite{Meyn_MarkovAggregation} proposed this cost function for deterministic aggregations, i.e., they proposed optimizing $\costpg$. Note that this restriction to deterministic aggregations comes at the loss of optimality: In~\cite[Example~3]{GeigerAmjad_HardClusters} a reversible, three-state Markov chain was given for which the optimal aggregation to $\card{\dom{Y}}=2$ states is stochastic. For the bi-partition problem, i.e., for $\card{\dom{Y}}=2$, the authors of~\cite{Meyn_MarkovAggregation} propose a relaxation to a spectral, i.e., eigenvector-based optimization problem, the solution of which has a computational complexity of $\mathcal{O}(\card{\dom{X}}^3)$. In general, this relaxation leads to a further loss of optimality, even among the search over all deterministic bi-partitions. For a general $\dom{Y}$, they suggest to solve the problem by repeated bi-partitioning, i.e., splitting sets of states until the desired cardinality is achieved.
	
	\item For $\beta=1$, the problem becomes equivalent to maximizing $\mutinf{X_1;Y_2}$. This is exactly the information bottleneck problem~\cite{Tishby_InformationBottleneck} for a Lagrangian parameter $\gamma\to\infty$:
	\begin{equation}\label{eq:IB}
	\mutinf{X_2;Y_2} - \gamma\mutinf{X_1;Y_2}.
	\end{equation}
	Algorithmic approaches to solving this optimization problem are introduced in~\cite{Slonim_PhD}. Note, that in this case the optimal aggregation will be deterministic~\cite[Thm.~1]{GeigerAmjad_HardClusters}. 
	\item For $\beta=0$, the authors of~\cite{GeigerEtAl_OptimalMarkovAggregation} relaxed their cost function $\ccostg{0}=\costlg$ as
	\begin{align}
	\ccostg{0} &= \ent{Y_2|Y_1}-\ent{Y_2|X_1} = \mutinf{Y_2;X_1|Y_1}\notag\\
	&\le \mutinf{X_2;X_1|Y_1} = \mutinf{X_2;X_1}-\mutinf{X_2;Y_1} \label{eq:relaxation}
	\end{align}
	and proposed using the agglomerative information bottleneck method~\cite{Slonim_Agglomerative} with the roles of $X_1$ and $X_2$ in~\eqref{eq:IB} exchanged to solve this relaxed optimization problem. The method has a computational complexity of $\mathcal{O}(\card{\dom{X}}^4)$~\cite[Sec.~3.4]{Slonim_PhD}. While the mapping minimizing $\costl$ may be stochastic, the mapping minimizing~\eqref{eq:relaxation} will be deterministic; hence, with this relaxation in mind, the restriction to deterministic aggregations made in~\cite{GeigerEtAl_OptimalMarkovAggregation} comes without an additional loss of optimality compared to what is lost in the relaxation.
	\item The authors of~\cite{Xu_Reduction} proposed minimizing 
	\begin{equation}\label{eq:XU_cost}
	\mutinf{X_1;X_2}  - \mutinf{X_2;Y_1} -\gamma \ent{Y_2|X_1}.
	\end{equation}
	They suggested using a deterministic annealing approach, reducing $\gamma$ successively until $\gamma=0$. In the limiting case, the cost function then coincides with~\eqref{eq:relaxation} and the optimal aggregation is again deterministic. Note that, for reversible Markov chains, we have $\mutinf{X_1;Y_2}=\mutinf{X_2;Y_1}$, hence both~\eqref{eq:relaxation} and~\eqref{eq:XU_cost} (for $\gamma=0$) are equivalent to $\ccost{1}$. Analyzing~\cite[Sec.~III.B]{Xu_Reduction} shows that in each annealing step the quantity
	\ifCLASSOPTIONdraftcls
		\begin{equation}
		 \kld{\pmf{X_2|X_1=x}}{\pmf{X_2|Y_1=y}}
		:= \sum_{x'\in\dom{X}} \pmf{X_2|X_1}(x'|x) \log \frac{\pmf{X_2|X_1}(x'|x)}{\pmf{X_2|Y_1}(x'|y)}
		\end{equation}
	\else
		\begin{multline}
		\kld{\pmf{X_2|X_1=x}}{\pmf{X_2|Y_1=y}} \\
		:= \sum_{x'\in\dom{X}} \pmf{X_2|X_1}(x'|x) \log \frac{\pmf{X_2|X_1}(x'|x)}{\pmf{X_2|Y_1}(x'|y)}
		\end{multline}
	\fi
	has to be computed for every $x$ and $y$. Hence, the computational complexity of this approach is $\mathcal{O}(\card{\mathcal{Y}}\cdot\card{\mathcal{X}}^2)$ in each annealing step.
\end{itemize}
 
\section{Markov Chain Aggregation Algorithms}\label{sec:algos}
We now propose an iterative method for optimizing \eqref{eq:finalcost} over deterministic aggregations for general values of $\beta$. The method consists of a sequential optimization algorithm (Algorithm~\ref{alg:sGITMA}) and an annealing procedure for $\beta$ (Algorithm~\ref{alg:Ann}) that prevents getting stuck in local optima. Since we focus only on deterministic aggregations, in the remainder of this section we can replace $\costbeta$ by $\costbetag$ for some $g{:}\ \dom{X}\to\dom{Y}$. Our algorithm has a computational complexity of $\mathcal{O}(\card{\mathcal{Y}}\cdot\card{\mathcal{X}}^2)$ per iteration. Note, however, that the restriction to deterministic aggregation functions comes, at least for some values of $\beta$, with a loss of optimality, i.e., in general we have $\min_\Wvec\  \costbeta \le \min_g\ \costbetag$.

\subsection{Sequential Algorithm}\label{subsec:seqalg}
We briefly illustrate an iteration of Algorithm~\ref{alg:sGITMA}: Suppose $x\in\mathcal{X}$ is mapped to the aggregate state $y\in\mathcal{Y}$, i.e., $g(x)=y$. We remove $x$ from aggregate state $y$. We then assign $x$ to every aggregate state $y'$, $y'\in\mathcal{Y}$, while keeping the rest of the mapping $g$ the same and evaluate the cost function. Finally, we assign $x$ to the aggregate state that minimized the cost function (breaking ties, if necessary). This procedure is repeated for every $x\in\mathcal{X}$. 

 \begin{algorithm}
 	\caption{Sequential Generalized Information-Theoretic Markov Aggregation.}\label{alg:sGITMA}
 	\begin{algorithmic}[1]
 		\Function {$g = $ sGITMA}{$\Pvec$, $\beta$, $\card{\mathcal{Y}}$, $\numitermax$, optional: initial aggregation function $g_{\text{init}}$}
 		\If {$g_{\text{init}}$ is empty} \Comment \textit{Initialization}
 		\State $g \gets $ Random Aggregation Function
 		\Else
 		\State $g \gets g_{\text{init}}$
 		\EndIf
 		\State $\numiter \gets 0$
 		\While {$\numiter < \numitermax$} \Comment \textit{Main Loop}
 		\For {all elements $x \in \mathcal{X}$} \Comment \textit{Optimizing $g$}
 		\For {all aggregate states $y \in \mathcal{Y}$}
 		\State $g_y(x') =
 		\begin{cases}
 		g(x')  &\quad x' \neq x   \\
 		y        &\quad x'=x		                          
 		\end{cases}$      \Comment \textit{Assign $x$ to aggregate state $y$}                             
 		\State $C_{g_y}= \cost_{\beta} (\Xvec,g_y)$
 		\EndFor
 		\State $g = \argmin\limits_{g_y} C_{g_y}$  \Comment \textit{(break ties)}    
 		\EndFor
 		\State $\numiter \gets \numiter + 1$
 		\EndWhile
 		\EndFunction
 	\end{algorithmic}
 \end{algorithm}

It is easy to verify that the cost function is reduced in each step of Algorithm~\ref{alg:sGITMA}, as a state is only assigned to a different aggregate state if the cost function is reduced. Hence, the algorithm modifies $g$ in each iteration in order to reduce the cost until it either reaches the maximum number of iterations or until the cost converges.

Note that the algorithm is random in the sense that it is started with a random aggregation function $g$. Depending on the specific application, though, a tailored initialization procedure may lead to performance improvements.

Finally, it is worth mentioning that for $\beta=1$ our Algorithm~\ref{alg:sGITMA} is equivalent to the sequential information bottleneck algorithm proposed in~\cite[Sec.~3.4]{Slonim_PhD}.

\subsection{Annealing Procedure for $\beta$}
Although Algorithm~\ref{alg:sGITMA} is guaranteed to converge (with proper tiebreaking), convergence to a global optimum is not ensured. The algorithm may get stuck in poor local minima. This happens particularly often for small values of $\beta$, as our experiments in Section~\ref{sec:experiments:QL} show. The reason is that, for small $\beta$, $\costbeta$ has many poor local minima and, randomly initialized, the algorithm is more likely to get stuck in one of them. In contrast, our results suggest that for larger values of $\beta$ the cost function has only few poor local minima and that the algorithm converges to a good local or a global minimum for a significant portion of random initializations.

A solution for small $\beta$ would thus be to choose an initialization that is close to a ``good'' local optimum. A simple idea is thus to re-use the function $g$ obtained for a large value of $\beta$ as initial aggregation for smaller values of $\beta$. We thus propose the following annealing algorithm: We initialize $\beta = 1$ to obtain $g$. Then, in each iteration of the annealing procedure, $\beta$ is reduced and the aggregation function is updated, starting from the result of the previous iteration. The procedure stops when $\beta$ reaches the desired value, $\beta_\text{target}$. The $\beta$-annealing algorithm is sketched as Algorithm \ref{alg:Ann}. As is clear from the description, the $\beta$-annealing algorithm closely follows graduated optimization in spirit \cite{blakegraduatedopt}. The results for  synthetic datasets with and without $\beta$-annealing are discussed in Section~\ref{sec:experiments:QL}, which show that without restarts one keeps getting stuck in bad local optima for small $\beta$, while with $\beta$-annealing one is able to avoid them. Furthermore in our experiments we have observed that $\beta$-annealing achieves good results for random initializations, hence tailoring initialization procedures is not necessary at least for the scenarios we considered.

 \begin{algorithm}
 	\caption{$\beta$-Annealing Information-Theoretic Markov Aggregation}\label{alg:Ann}
 	\begin{algorithmic}[1]
 		\Function {$g =$ AnnITMA}{$\Pvec$, $\beta_\text{target}$, $\card{\mathcal{Y}}$, $\numitermax$, $\Delta$}
 		\State $\beta \gets 1$
 		\State $g$ = sGITMA($\Pvec$, $\beta$, $\card{\mathcal{Y}}$, $\numitermax$) \Comment \textit{Inizialization}
 		\While {$\beta > \betatar$}
 		\State $\beta \gets \max\{\beta - \Delta,\betatar\}$ 
 		\State $g$ = sGITMA($\Pvec$, $\beta$, $\card{\mathcal{Y}}$, $\numitermax$, $g$)
 		\EndWhile
 		\EndFunction
 	\end{algorithmic}
 \end{algorithm}

Note that the $\beta$-annealing algorithm admits producing results for a series of values of $\beta$ at once: Keeping all intermediate aggregation functions, one obtains aggregations for all values of $\beta$ in the set $\{1, 1-\Delta,1-2\Delta,\dots,1-\Delta\lceil\frac{1-\betatar}{\Delta}\rceil,\betatar\}$. The aggregations one obtains are exactly those one would obtain from restarting \textsc{AnnITMA} for each value in this set, each time with the same random initial partition. We used this fact in our experiments: If we were interested in results for $\betatar$ ranging between 0 and 1 in steps of 0.05, rather than restarting \textsc{AnnITMA} for each value in this set, we started \textsc{AnnITMA} for $\betatar=0$ and $\Delta=0.05$ once, keeping all intermediate results.

\subsection{Computational Complexity of the Sequential Algorithm}
Note that the asymptotic computational complexity of Algorithm~\ref{alg:Ann} equals that of Algorithm~\ref{alg:sGITMA}, since the former  simply calls the latter $\lceil (1-\betatar)/\Delta\rceil+1$ times. We thus only evaluate the complexity of Algorithm~\ref{alg:sGITMA}. To this end, we first observe that the cost $\cost_{\beta} (\Xvec,\Wvec)$ can be expressed with only three mutual information terms for any $\beta$ and $\Wvec$:
\ifCLASSOPTIONdraftcls
	\begin{equation}
	\costbetag = \beta\mutinf{X_1;X_2} +(1-2\beta)\mutinf{X_1;g(X_2)} - (1-\beta)\mutinf{g(X_1);g(X_2)}.
	\end{equation}
\else
	\begin{multline}
	\costbetag = \beta\mutinf{X_1;X_2} +(1-2\beta)\mutinf{X_1;g(X_2)} \\- (1-\beta)\mutinf{g(X_1);g(X_2)}.
	\end{multline} 
\fi
The first term $\mutinf{X_1;X_2}$ is constant regardless of the aggregation hence the computation of $\costbetag$ depends upon the computation of the other two terms.

In each iteration of the main loop, we evaluate $\cost_{\beta} (\Xvec,g_y)$ in line 12 for each $x \in \mathcal{X}$ and $y \in \mathcal{Y}$. Note that $g_y$ differs from the current $g$ only for one element as defined in line 11. Thus, the joint PMF $\pmf{X_1, g_y(X_2)}$ differs from $\pmf{X_1, g(X_2)}$ in only two rows and hence can be computed from $\pmf{X_1, g(X_2)}$ in $\mathcal{O}(\card{\mathcal{X}})$ computations. Moreover, $\mutinf{X_1;g_y(X_2)}$ can be computed from $\mutinf{X_1;g(X_2)}$ in $\mathcal{O}(\card{\mathcal{X}})$ computations, cf.~\cite[Prop.~1]{Slonim_Agglomerative}. This is due to the fact that we can write~\cite[eq.~(2.28)]{Cover_Information1} 

\begin{equation}
\begin{split}
&\mutinf{X_1;g_y(X_2)} = \mutinf{X_1;g(X_2)}   \\
&+ \sum_{\substack{x_1 \in \mathcal{X} \\  y_2\in\{y,g(x)\}} } 
\pmf{X_1, g_y(X_2)}(x_1, y_2) 
\log \left( \frac{\pmf{X_1, g_y(X_2)}(x_1, y_2)}{\pmf{X_1}(x_1) \pmf{g_y(X_2)}(y_2)} \right) \\
& - \sum_{\substack{x_1 \in \mathcal{X} \\  y_2\in\{y,g(x)\}} } 
\pmf{X_1, g(X_2)}(x_1, y_2)
\log \left( \frac{\pmf{X_1, g(X_2)}(x_1, y_2)}{\pmf{X_1}(x_1) \pmf{g(X_2)}(y_2)} \right).
\end{split}
\end{equation}
The term $\mutinf{g_y(X_1);g_y(X_2)}$ can be computed from $\mutinf{g(X_1);g(X_2)}$ in $\mathcal{O}(\card{\mathcal{Y}})$ computations, but requires the updated joint PMF $\pmf{g_y(X_1),g_y(X_2)}$. This PMF can be computed from $\pmf{g(X_1),g(X_2)}$ in $\mathcal{O}(\card{\mathcal{X}})$ computations. Combining this with the fact that line 12 is executed once for each aggregate state in $\mathcal{Y}$ and once for each state in $\mathcal{X}$ in every iteration, we get that optimizing $g$ has a computational complexity of $\mathcal{O}(\card{\mathcal{Y}}\cdot\card{\mathcal{X}}^2)$ per iteration.

\begin{figure*}[t]
\def \figheight {1.7in}
\def \figwidth {1.5in}
\subfigure[$\alpha=\varepsilon=0$]{\includegraphics[width=0.25\textwidth]{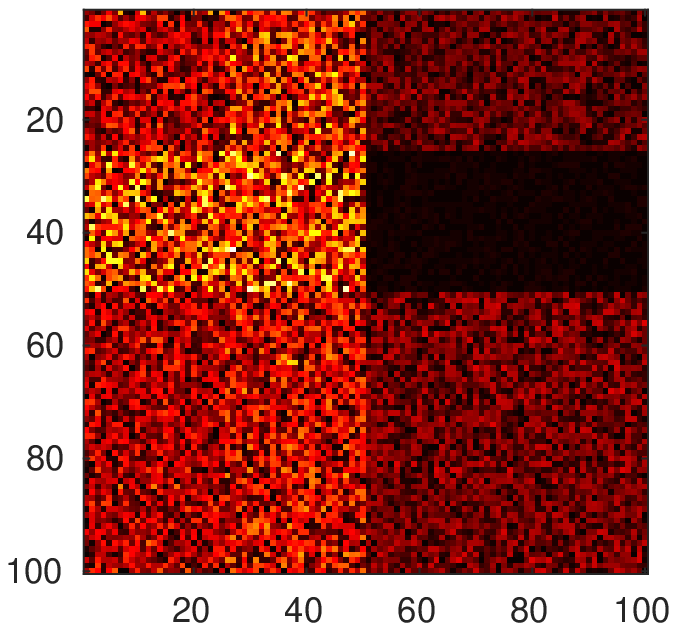}}\hfill
\subfigure[$\alpha=0.95$, $\varepsilon=0.8$]{\includegraphics[width=0.25\textwidth]{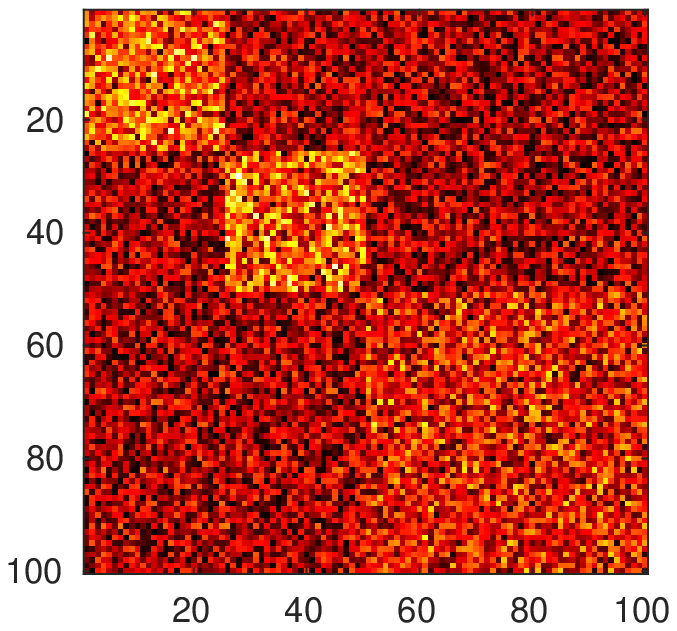}}\hfill
\subfigure[$\alpha=0.95$, $\varepsilon=0.4$]{\includegraphics[width=0.25\textwidth]{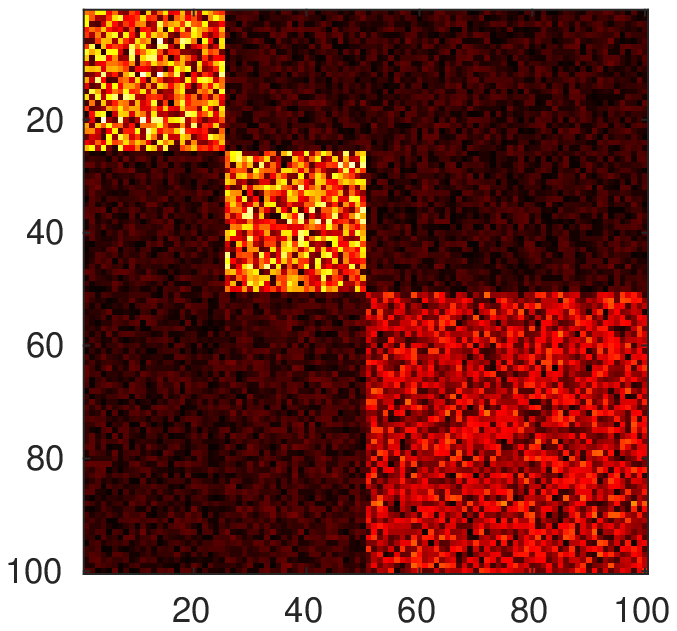}}\hfill
\subfigure[$\alpha=0.95$, $\varepsilon=0.4$; rows and columns are permuted\label{fig:results:perm}]{\includegraphics[width=0.24\textwidth]{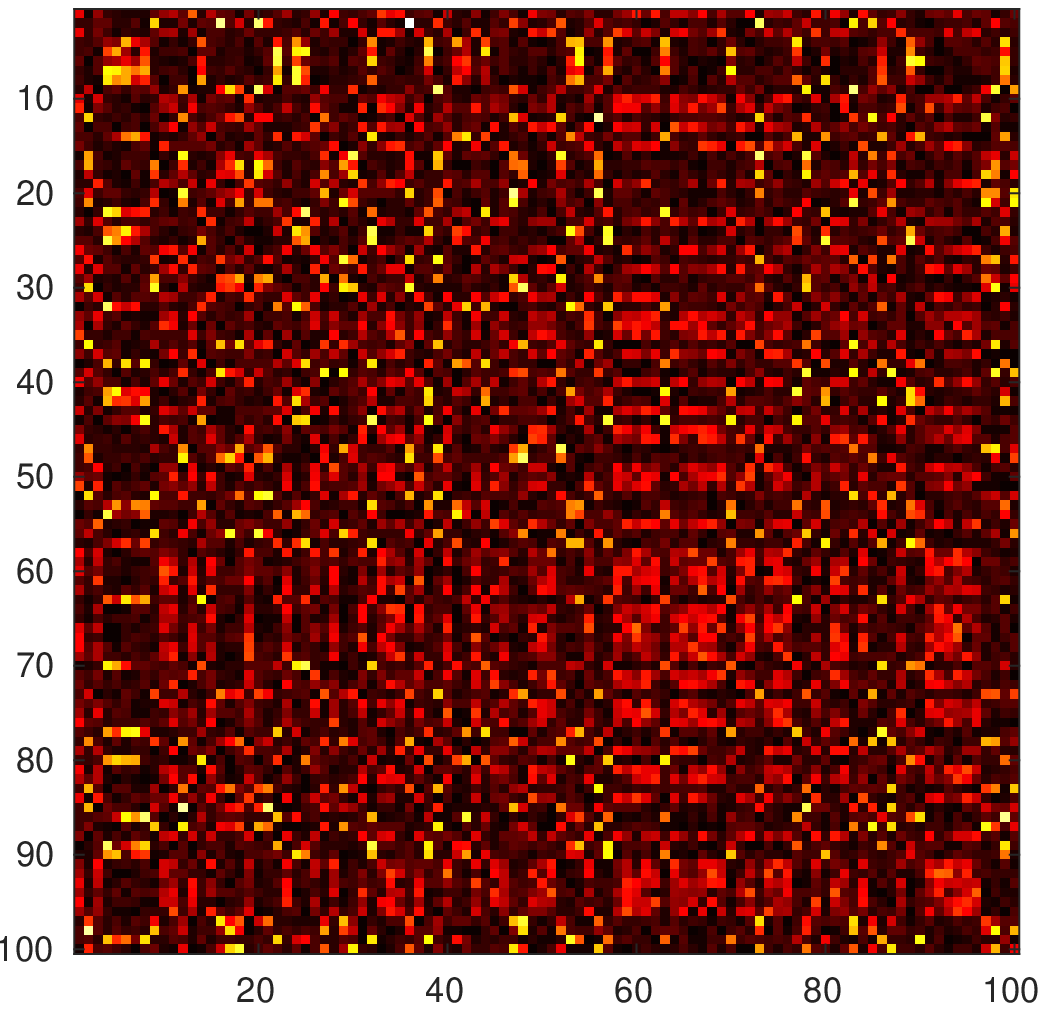}}\\
\subfigure[Cost with $\beta$-annealing, $\varepsilon=0.4$\label{fig:results:cost}]{
%
%
\begin{tikzpicture}

\begin{axis}[%
width=\figheight,
height=\figwidth,
at={(0.686in,0.48in)},
scale only axis,
xmin=0,
xmax=1,
xlabel style={font=\color{white!15!black}},
xlabel={$\beta$},
ymin=0,
ymax=0.2,
ylabel style={font=\color{white!15!black},at={(0.05,0.5)}},
ylabel={$\text{C}_\beta$},
axis background/.style={fill=white},
legend style={at={(0.97,0.03)}, anchor=south east, legend cell align=left, align=left, draw=white!15!black}
]

\addplot[area legend, draw=none, fill=blue, fill opacity=0.15, forget plot]
table[row sep=crcr] {%
x	y\\
1	0.139239970745593\\
0.9	0.125714306998743\\
0.8	0.111886490304619\\
0.7	0.0978417923317633\\
0.6	0.0836353584392778\\
0.5	0.0700961427735121\\
0.4	0.0559389834170275\\
0.3	0.0422590983901165\\
0.2	0.0282655332653578\\
0.1	0.0142179793803353\\
0	0.000131678955660925\\
0	0.00172041936850214\\
0.1	0.0171142072742457\\
0.2	0.0323995212656888\\
0.3	0.0477998251041784\\
0.4	0.0637820346316966\\
0.5	0.0787785243966338\\
0.6	0.0951176595467724\\
0.7	0.109821350354659\\
0.8	0.124634457246493\\
0.9	0.13947092479137\\
1	0.154192993270982\\
}--cycle;
\addplot [color=white!55!blue, forget plot]
  table[row sep=crcr]{%
1	0.139239970745593\\
0.9	0.125714306998743\\
0.8	0.111886490304619\\
0.7	0.0978417923317633\\
0.6	0.0836353584392778\\
0.5	0.0700961427735121\\
0.4	0.0559389834170275\\
0.3	0.0422590983901165\\
0.2	0.0282655332653578\\
0.1	0.0142179793803353\\
0	0.000131678955660925\\
};
\addplot [color=white!55!blue, forget plot]
  table[row sep=crcr]{%
1	0.154192993270982\\
0.9	0.13947092479137\\
0.8	0.124634457246493\\
0.7	0.109821350354659\\
0.6	0.0951176595467724\\
0.5	0.0787785243966338\\
0.4	0.0637820346316966\\
0.3	0.0477998251041784\\
0.2	0.0323995212656888\\
0.1	0.0171142072742457\\
0	0.00172041936850214\\
};
\addplot [color=blue]
  table[row sep=crcr]{%
1	0.146716482008288\\
0.9	0.132592615895056\\
0.8	0.118260473775556\\
0.7	0.103831571343211\\
0.6	0.0893765089930251\\
0.5	0.074437333585073\\
0.4	0.0598605090243621\\
0.3	0.0450294617471475\\
0.2	0.0303325272655233\\
0.1	0.0156660933272905\\
0	0.000926049162081531\\
};

\addplot[area legend, draw=none, fill=red, fill opacity=0.15, forget plot]
table[row sep=crcr] {%
x	y\\
1	0.144471509455951\\
0.9	0.130069103611749\\
0.8	0.115668415219256\\
0.7	0.101262465662939\\
0.6	0.0868597567602896\\
0.5	0.0724738536089577\\
0.4	0.0580925101036231\\
0.3	0.0437144730931751\\
0.2	0.0293265875027867\\
0.1	0.0149064748990931\\
0	0.000276806729076262\\
0	0.00172456775276706\\
0.1	0.0177800073805595\\
0.2	0.0340426008716253\\
0.3	0.0503286131728527\\
0.4	0.0666206716202923\\
0.5	0.0829009957424652\\
0.6	0.099168705268339\\
0.7	0.11539836064366\\
0.8	0.131620148838759\\
0.9	0.147847275963686\\
1	0.164067192477363\\
}--cycle;
\addplot [color=white!55!red, forget plot]
  table[row sep=crcr]{%
1	0.144471509455951\\
0.9	0.130069103611749\\
0.8	0.115668415219256\\
0.7	0.101262465662939\\
0.6	0.0868597567602896\\
0.5	0.0724738536089577\\
0.4	0.0580925101036231\\
0.3	0.0437144730931751\\
0.2	0.0293265875027867\\
0.1	0.0149064748990931\\
0	0.000276806729076262\\
};
\addplot [color=white!55!red, forget plot]
  table[row sep=crcr]{%
1	0.164067192477363\\
0.9	0.147847275963686\\
0.8	0.131620148838759\\
0.7	0.11539836064366\\
0.6	0.099168705268339\\
0.5	0.0829009957424652\\
0.4	0.0666206716202923\\
0.3	0.0503286131728527\\
0.2	0.0340426008716253\\
0.1	0.0177800073805595\\
0	0.00172456775276706\\
};
\addplot [color=red]
  table[row sep=crcr]{%
1	0.154269350966657\\
0.9	0.138958189787718\\
0.8	0.123644282029007\\
0.7	0.108330413153299\\
0.6	0.0930142310143143\\
0.5	0.0776874246757115\\
0.4	0.0623565908619577\\
0.3	0.0470215431330139\\
0.2	0.031684594187206\\
0.1	0.0163432411398263\\
0	0.00100068724092166\\
};

\addplot[area legend, draw=none, fill=green, fill opacity=0.15, forget plot]
table[row sep=crcr] {%
x	y\\
1	0.186292413024049\\
0.9	0.167793890656289\\
0.8	0.149295302904992\\
0.7	0.130796621796286\\
0.6	0.112297800760172\\
0.5	0.0937987561210882\\
0.4	0.0752993210398182\\
0.3	0.0567991085108004\\
0.2	0.0382969735578284\\
0.1	0.0197874829001345\\
0	0.00117766138191576\\
0	0.00144436798165785\\
0.1	0.0205855087685656\\
0.2	0.0398269804159981\\
0.3	0.0590758077681528\\
0.4	0.0783265575442614\\
0.5	0.0975780847681179\\
0.6	0.116830002434161\\
0.7	0.136082143703173\\
0.8	0.155334424899593\\
0.9	0.174586799453423\\
1	0.193839239390789\\
}--cycle;
\addplot [color=white!55!green, forget plot]
  table[row sep=crcr]{%
1	0.186292413024049\\
0.9	0.167793890656289\\
0.8	0.149295302904992\\
0.7	0.130796621796286\\
0.6	0.112297800760172\\
0.5	0.0937987561210882\\
0.4	0.0752993210398182\\
0.3	0.0567991085108004\\
0.2	0.0382969735578284\\
0.1	0.0197874829001345\\
0	0.00117766138191576\\
};
\addplot [color=white!55!green, forget plot]
  table[row sep=crcr]{%
1	0.193839239390789\\
0.9	0.174586799453423\\
0.8	0.155334424899593\\
0.7	0.136082143703173\\
0.6	0.116830002434161\\
0.5	0.0975780847681179\\
0.4	0.0783265575442614\\
0.3	0.0590758077681528\\
0.2	0.0398269804159981\\
0.1	0.0205855087685656\\
0	0.00144436798165785\\
};
\addplot [color=green]
  table[row sep=crcr]{%
1	0.190065826207419\\
0.9	0.171190345054856\\
0.8	0.152314863902293\\
0.7	0.133439382749729\\
0.6	0.114563901597166\\
0.5	0.0956884204446031\\
0.4	0.0768129392920398\\
0.3	0.0579374581394766\\
0.2	0.0390619769869133\\
0.1	0.0201864958343501\\
0	0.00131101468178681\\
};

\end{axis}
\end{tikzpicture}%
}\hfill
\subfigure[Cost without $\beta$-annealing, $\varepsilon=0.4$\label{fig:results:costseq}]{
%
%
\begin{tikzpicture}

\begin{axis}[%
width=\figheight,
height=\figwidth,
at={(0.686in,0.48in)},
scale only axis,
xmin=0,
xmax=1,
xlabel style={font=\color{white!15!black}},
xlabel={$\beta$},
ymin=0,
ymax=0.25,
ylabel style={font=\color{white!15!black},at={(0.05,0.5)}},
ylabel={$\text{C}_\beta$},
axis background/.style={fill=white},
]

\addplot[area legend, draw=none, fill=blue, fill opacity=0.15, forget plot]
table[row sep=crcr] {%
x	y\\
1	0.139429378823514\\
0.9	0.125439470472263\\
0.8	0.111596089047858\\
0.7	0.0984339034082787\\
0.6	0.0842617287220652\\
0.5	0.0701965146759753\\
0.4	0.0583977208846867\\
0.3	0.0490454803030899\\
0.2	0.0330233851671212\\
0.1	0.0169133111051406\\
0	0.00076630279128209\\
0	0.00119009756576304\\
0.1	0.0240746477135047\\
0.2	0.0469882182528285\\
0.3	0.069835075898439\\
0.4	0.0890517587663541\\
0.5	0.0786600559356249\\
0.6	0.0937789890991887\\
0.7	0.108862052869518\\
0.8	0.124721841976429\\
0.9	0.140382858312426\\
1	0.153616981291397\\
}--cycle;
\addplot [color=white!55!blue, forget plot]
  table[row sep=crcr]{%
1	0.139429378823514\\
0.9	0.125439470472263\\
0.8	0.111596089047858\\
0.7	0.0984339034082787\\
0.6	0.0842617287220652\\
0.5	0.0701965146759753\\
0.4	0.0583977208846867\\
0.3	0.0490454803030899\\
0.2	0.0330233851671212\\
0.1	0.0169133111051406\\
0	0.00076630279128209\\
};
\addplot [color=white!55!blue, forget plot]
  table[row sep=crcr]{%
1	0.153616981291397\\
0.9	0.140382858312426\\
0.8	0.124721841976429\\
0.7	0.108862052869518\\
0.6	0.0937789890991887\\
0.5	0.0786600559356249\\
0.4	0.0890517587663541\\
0.3	0.069835075898439\\
0.2	0.0469882182528285\\
0.1	0.0240746477135047\\
0	0.00119009756576304\\
};
\addplot [color=blue]
  table[row sep=crcr]{%
1	0.146523180057455\\
0.9	0.132911164392345\\
0.8	0.118158965512144\\
0.7	0.103647978138898\\
0.6	0.0890203589106269\\
0.5	0.0744282853058001\\
0.4	0.0737247398255204\\
0.3	0.0594402781007644\\
0.2	0.0400058017099749\\
0.1	0.0204939794093227\\
0	0.000978200178522564\\
};

\addplot[area legend, draw=none, fill=red, fill opacity=0.15, forget plot]
table[row sep=crcr] {%
x	y\\
1	0.143332205369124\\
0.9	0.130163255427594\\
0.8	0.115818789028204\\
0.7	0.100555900229212\\
0.6	0.0862764642699184\\
0.5	0.0693103422257394\\
0.4	0.0939662734161734\\
0.3	0.0743123971415783\\
0.2	0.0499600934151186\\
0.1	0.025551216202843\\
0	0.00108330380238298\\
0	0.00167558492029128\\
0.1	0.0343129045618239\\
0.2	0.066999266205605\\
0.3	0.0996907630924947\\
0.4	0.131848106489528\\
0.5	0.089272597242259\\
0.6	0.100302395877323\\
0.7	0.117221018511724\\
0.8	0.131480326623522\\
0.9	0.148858703424625\\
1	0.167977891222437\\
}--cycle;
\addplot [color=white!55!red, forget plot]
  table[row sep=crcr]{%
1	0.143332205369124\\
0.9	0.130163255427594\\
0.8	0.115818789028204\\
0.7	0.100555900229212\\
0.6	0.0862764642699184\\
0.5	0.0693103422257394\\
0.4	0.0939662734161734\\
0.3	0.0743123971415783\\
0.2	0.0499600934151186\\
0.1	0.025551216202843\\
0	0.00108330380238298\\
};
\addplot [color=white!55!red, forget plot]
  table[row sep=crcr]{%
1	0.167977891222437\\
0.9	0.148858703424625\\
0.8	0.131480326623522\\
0.7	0.117221018511724\\
0.6	0.100302395877323\\
0.5	0.089272597242259\\
0.4	0.131848106489528\\
0.3	0.0996907630924947\\
0.2	0.066999266205605\\
0.1	0.0343129045618239\\
0	0.00167558492029128\\
};
\addplot [color=red]
  table[row sep=crcr]{%
1	0.155655048295781\\
0.9	0.13951097942611\\
0.8	0.123649557825863\\
0.7	0.108888459370468\\
0.6	0.0932894300736205\\
0.5	0.0792914697339992\\
0.4	0.112907189952851\\
0.3	0.0870015801170365\\
0.2	0.0584796798103618\\
0.1	0.0299320603823334\\
0	0.00137944436133713\\
};

\addplot[area legend, draw=none, fill=green, fill opacity=0.15, forget plot]
table[row sep=crcr] {%
x	y\\
1	0.176274809644309\\
0.9	0.161768466740437\\
0.8	0.141014569282157\\
0.7	0.123662044092665\\
0.6	0.106431738468237\\
0.5	0.0822573331694121\\
0.4	0.239923402171356\\
0.3	0.180997773367848\\
0.2	0.121634249998083\\
0.1	0.0621710026939598\\
0	0.00249634642217859\\
0	0.0033430056571576\\
0.1	0.0641961680431241\\
0.2	0.12542422209672\\
0.3	0.186582876394007\\
0.4	0.247372766886974\\
0.5	0.114396923584996\\
0.6	0.124240416371033\\
0.7	0.145033634370173\\
0.8	0.165780230400921\\
0.9	0.181817581540921\\
1	0.206547204791923\\
}--cycle;
\addplot [color=white!55!green, forget plot]
  table[row sep=crcr]{%
1	0.176274809644309\\
0.9	0.161768466740437\\
0.8	0.141014569282157\\
0.7	0.123662044092665\\
0.6	0.106431738468237\\
0.5	0.0822573331694121\\
0.4	0.239923402171356\\
0.3	0.180997773367848\\
0.2	0.121634249998083\\
0.1	0.0621710026939598\\
0	0.00249634642217859\\
};
\addplot [color=white!55!green, forget plot]
  table[row sep=crcr]{%
1	0.206547204791923\\
0.9	0.181817581540921\\
0.8	0.165780230400921\\
0.7	0.145033634370173\\
0.6	0.124240416371033\\
0.5	0.114396923584996\\
0.4	0.247372766886974\\
0.3	0.186582876394007\\
0.2	0.12542422209672\\
0.1	0.0641961680431241\\
0	0.0033430056571576\\
};
\addplot [color=green]
  table[row sep=crcr]{%
1	0.191411007218116\\
0.9	0.171793024140679\\
0.8	0.153397399841539\\
0.7	0.134347839231419\\
0.6	0.115336077419635\\
0.5	0.098327128377204\\
0.4	0.243648084529165\\
0.3	0.183790324880928\\
0.2	0.123529236047401\\
0.1	0.063183585368542\\
0	0.0029196760396681\\
};

\end{axis}
\end{tikzpicture}%
}\hfill
\subfigure[ARI without $\beta$-annealing, $\varepsilon=0.4$\label{fig:results:ARIseq}]{
%
%
\begin{tikzpicture}

\begin{axis}[%
width=\figheight,
height=\figwidth,
at={(0.603in,0.48in)},
scale only axis,
xmin=0,
xmax=1,
xlabel style={font=\color{white!15!black}},
xlabel={$\beta$},
ymin=-0.1,
ymax=1.2,
ylabel style={font=\color{white!15!black},at={(0.1,0.5)}},
ylabel={ARI},
axis background/.style={fill=white},
]

\addplot[area legend, draw=none, fill=blue, fill opacity=0.15, forget plot]
table[row sep=crcr] {%
x	y\\
1	0.51329628952956\\
0.9	0.603276039127817\\
0.8	0.670578446065537\\
0.7	0.694660391590727\\
0.6	0.734034391216106\\
0.5	0.729292247740697\\
0.4	-0.157870270451251\\
0.3	-0.0306535617685319\\
0.2	-0.0237857177486148\\
0.1	-0.0230290469238914\\
0	-0.023378251684329\\
0	0.004854419229101\\
0.1	0.00612421920916531\\
0.2	0.0079280905759374\\
0.3	0.022855968280045\\
0.4	0.72391230935884\\
0.5	1.1322686616933\\
0.6	1.11753384160333\\
0.7	1.11495492104435\\
0.8	1.11130936878523\\
0.9	1.08361243699102\\
1	1.03164471141187\\
}--cycle;
\addplot [color=white!55!blue, forget plot]
  table[row sep=crcr]{%
1	0.51329628952956\\
0.9	0.603276039127817\\
0.8	0.670578446065537\\
0.7	0.694660391590727\\
0.6	0.734034391216106\\
0.5	0.729292247740697\\
0.4	-0.157870270451251\\
0.3	-0.0306535617685319\\
0.2	-0.0237857177486148\\
0.1	-0.0230290469238914\\
0	-0.023378251684329\\
};
\addplot [color=white!55!blue, forget plot]
  table[row sep=crcr]{%
1	1.03164471141187\\
0.9	1.08361243699102\\
0.8	1.11130936878523\\
0.7	1.11495492104435\\
0.6	1.11753384160333\\
0.5	1.1322686616933\\
0.4	0.72391230935884\\
0.3	0.022855968280045\\
0.2	0.0079280905759374\\
0.1	0.00612421920916531\\
0	0.004854419229101\\
};
\addplot [color=blue]
  table[row sep=crcr]{%
1	0.772470500470717\\
0.9	0.843444238059417\\
0.8	0.890943907425381\\
0.7	0.904807656317537\\
0.6	0.925784116409718\\
0.5	0.930780454717\\
0.4	0.283021019453794\\
0.3	-0.00389879674424346\\
0.2	-0.00792881358633873\\
0.1	-0.00845241385736306\\
0	-0.00926191622761402\\
};

\addplot[area legend, draw=none, fill=red, fill opacity=0.15, forget plot]
table[row sep=crcr] {%
x	y\\
1	0.796258052814338\\
0.9	0.818254319512593\\
0.8	0.852947984453524\\
0.7	0.829956704120078\\
0.6	0.841102468278426\\
0.5	0.794412858147392\\
0.4	-0.160106379254789\\
0.3	-0.0191919021779187\\
0.2	-0.0191924166643048\\
0.1	-0.0190372573419281\\
0	-0.0191771169010306\\
0	-0.0162248967204724\\
0.1	-0.0163722936628258\\
0.2	-0.0161614428879553\\
0.3	-0.0158638712050855\\
0.4	0.246020825897032\\
0.5	1.11314315739513\\
0.6	1.09491773898913\\
0.7	1.09787044889451\\
0.8	1.09093674892823\\
0.9	1.10031296996906\\
1	1.10406454081591\\
}--cycle;
\addplot [color=white!55!red, forget plot]
  table[row sep=crcr]{%
1	0.796258052814338\\
0.9	0.818254319512593\\
0.8	0.852947984453524\\
0.7	0.829956704120078\\
0.6	0.841102468278426\\
0.5	0.794412858147392\\
0.4	-0.160106379254789\\
0.3	-0.0191919021779187\\
0.2	-0.0191924166643048\\
0.1	-0.0190372573419281\\
0	-0.0191771169010306\\
};
\addplot [color=white!55!red, forget plot]
  table[row sep=crcr]{%
1	1.10406454081591\\
0.9	1.10031296996906\\
0.8	1.09093674892823\\
0.7	1.09787044889451\\
0.6	1.09491773898913\\
0.5	1.11314315739513\\
0.4	0.246020825897032\\
0.3	-0.0158638712050855\\
0.2	-0.0161614428879553\\
0.1	-0.0163722936628258\\
0	-0.0162248967204724\\
};
\addplot [color=red]
  table[row sep=crcr]{%
1	0.950161296815123\\
0.9	0.959283644740829\\
0.8	0.971942366690875\\
0.7	0.963913576507292\\
0.6	0.968010103633777\\
0.5	0.953778007771259\\
0.4	0.0429572233211213\\
0.3	-0.0175278866915021\\
0.2	-0.01767692977613\\
0.1	-0.0177047755023769\\
0	-0.0177010068107515\\
};

\addplot[area legend, draw=none, fill=green, fill opacity=0.15, forget plot]
table[row sep=crcr] {%
x	y\\
1	0.947430992426494\\
0.9	0.964352728817666\\
0.8	0.952361503841091\\
0.7	0.947860417168497\\
0.6	0.948440334876856\\
0.5	0.887725325476174\\
0.4	-0.0188616060346536\\
0.3	-0.0190152606496481\\
0.2	-0.0189374470387012\\
0.1	-0.0189338771343465\\
0	-0.0189433099535263\\
0	-0.017153524101874\\
0.1	-0.017168279757719\\
0.2	-0.0172471512150083\\
0.3	-0.0170051159800747\\
0.4	-0.0171686528349943\\
0.5	1.0798488289075\\
0.6	1.04307796913363\\
0.7	1.04356297474804\\
0.8	1.03984502701147\\
0.9	1.03140642318758\\
1	1.04392127218493\\
}--cycle;
\addplot [color=white!55!green, forget plot]
  table[row sep=crcr]{%
1	0.947430992426494\\
0.9	0.964352728817666\\
0.8	0.952361503841091\\
0.7	0.947860417168497\\
0.6	0.948440334876856\\
0.5	0.887725325476174\\
0.4	-0.0188616060346536\\
0.3	-0.0190152606496481\\
0.2	-0.0189374470387012\\
0.1	-0.0189338771343465\\
0	-0.0189433099535263\\
};
\addplot [color=white!55!green, forget plot]
  table[row sep=crcr]{%
1	1.04392127218493\\
0.9	1.03140642318758\\
0.8	1.03984502701147\\
0.7	1.04356297474804\\
0.6	1.04307796913363\\
0.5	1.0798488289075\\
0.4	-0.0171686528349943\\
0.3	-0.0170051159800747\\
0.2	-0.0172471512150083\\
0.1	-0.017168279757719\\
0	-0.017153524101874\\
};
\addplot [color=green]
  table[row sep=crcr]{%
1	0.99567613230571\\
0.9	0.997879576002623\\
0.8	0.996103265426282\\
0.7	0.995711695958266\\
0.6	0.995759152005245\\
0.5	0.983787077191836\\
0.4	-0.018015129434824\\
0.3	-0.0180101883148614\\
0.2	-0.0180922991268548\\
0.1	-0.0180510784460328\\
0	-0.0180484170277002\\
};

\end{axis}
\end{tikzpicture}%
}\\
\subfigure[ARI with $\beta$-annealing, $\varepsilon=0$\label{fig:results:ARI0}]{
%
%
\begin{tikzpicture}

\begin{axis}[%
width=\figheight,
height=\figwidth,
at={(0.552in,0.48in)},
scale only axis,
xmin=0,
xmax=1,
xlabel style={font=\color{white!15!black}},
xlabel={$\beta$},
ymin=-0.1,
ymax=1.2,
ylabel style={font=\color{white!15!black},at={(0.1,0.5)}},
ylabel={ARI},
axis background/.style={fill=white},
legend style={at={(0.97,0.03)}, anchor=south east, legend cell align=left, align=left, draw=white!15!black}
]

\addplot[area legend, draw=none, fill=blue, fill opacity=0.15, forget plot]
table[row sep=crcr] {%
x	y\\
1	0.524934521441569\\
0.9	0.603076481428833\\
0.8	0.650397995785699\\
0.7	0.698455300047955\\
0.6	0.71996765217518\\
0.5	0.774210450193827\\
0.4	0.791120275792699\\
0.3	0.795718945299139\\
0.2	0.802084491679333\\
0.1	0.802302504373299\\
0	0.801008421016873\\
0	1.10669934926812\\
0.1	1.10601518794628\\
0.2	1.10612432053413\\
0.3	1.10767009211092\\
0.4	1.10822589164732\\
0.5	1.10927269123097\\
0.6	1.11014668495371\\
0.7	1.10588805251526\\
0.8	1.09634706392412\\
0.9	1.07789082021279\\
1	1.03697083633297\\
}--cycle;
\addplot [color=white!55!blue, forget plot]
  table[row sep=crcr]{%
1	0.524934521441569\\
0.9	0.603076481428833\\
0.8	0.650397995785699\\
0.7	0.698455300047955\\
0.6	0.71996765217518\\
0.5	0.774210450193827\\
0.4	0.791120275792699\\
0.3	0.795718945299139\\
0.2	0.802084491679333\\
0.1	0.802302504373299\\
0	0.801008421016873\\
};
\addplot [color=white!55!blue, forget plot]
  table[row sep=crcr]{%
1	1.03697083633297\\
0.9	1.07789082021279\\
0.8	1.09634706392412\\
0.7	1.10588805251526\\
0.6	1.11014668495371\\
0.5	1.10927269123097\\
0.4	1.10822589164732\\
0.3	1.10767009211092\\
0.2	1.10612432053413\\
0.1	1.10601518794628\\
0	1.10669934926812\\
};
\addplot [color=blue]
  table[row sep=crcr]{%
1	0.780952678887272\\
0.9	0.840483650820809\\
0.8	0.873372529854912\\
0.7	0.902171676281609\\
0.6	0.915057168564447\\
0.5	0.941741570712396\\
0.4	0.949673083720009\\
0.3	0.951694518705031\\
0.2	0.954104406106732\\
0.1	0.954158846159791\\
0	0.953853885142496\\
};
\addlegendentry{$\alpha\text{=0 (lumpable)}$}

\addplot[area legend, draw=none, fill=red, fill opacity=0.15, forget plot]
table[row sep=crcr] {%
x	y\\
1	0.829661270561596\\
0.9	0.829730052837793\\
0.8	0.829757799293711\\
0.7	0.829792769492597\\
0.6	0.829334429024198\\
0.5	0.827892420960586\\
0.4	0.828905197802876\\
0.3	0.829034158643073\\
0.2	0.828204230156227\\
0.1	0.816615254004916\\
0	0.816886763648912\\
0	1.10668136780859\\
0.1	1.10682734457191\\
0.2	1.09884836329908\\
0.3	1.09837718770931\\
0.4	1.09844790088486\\
0.5	1.09902375630467\\
0.6	1.0982242356829\\
0.7	1.09795657966578\\
0.8	1.09797665237433\\
0.9	1.09799257136586\\
1	1.09789125576828\\
}--cycle;
\addplot [color=white!55!red, forget plot]
  table[row sep=crcr]{%
1	0.829661270561596\\
0.9	0.829730052837793\\
0.8	0.829757799293711\\
0.7	0.829792769492597\\
0.6	0.829334429024198\\
0.5	0.827892420960586\\
0.4	0.828905197802876\\
0.3	0.829034158643073\\
0.2	0.828204230156227\\
0.1	0.816615254004916\\
0	0.816886763648912\\
};
\addplot [color=white!55!red, forget plot]
  table[row sep=crcr]{%
1	1.09789125576828\\
0.9	1.09799257136586\\
0.8	1.09797665237433\\
0.7	1.09795657966578\\
0.6	1.0982242356829\\
0.5	1.09902375630467\\
0.4	1.09844790088486\\
0.3	1.09837718770931\\
0.2	1.09884836329908\\
0.1	1.10682734457191\\
0	1.10668136780859\\
};
\addplot [color=red]
  table[row sep=crcr]{%
1	0.963776263164937\\
0.9	0.963861312101825\\
0.8	0.963867225834021\\
0.7	0.963874674579188\\
0.6	0.96377933235355\\
0.5	0.96345808863263\\
0.4	0.96367654934387\\
0.3	0.963705673176193\\
0.2	0.963526296727655\\
0.1	0.961721299288411\\
0	0.96178406572875\\
};
\addlegendentry{$\alpha\text{=0.5}$}

\addplot[area legend, draw=none, fill=green, fill opacity=0.15, forget plot]
table[row sep=crcr] {%
x	y\\
1	0.963521590317077\\
0.9	0.963521590317077\\
0.8	0.963521590317077\\
0.7	0.963521590317077\\
0.6	0.963554926784675\\
0.5	0.963654890614917\\
0.4	0.964352728817666\\
0.3	0.964352728817666\\
0.2	0.963821345245808\\
0.1	0.963821345245808\\
0	0.963821345245808\\
0	1.0318745896637\\
0.1	1.0318745896637\\
0.2	1.0318745896637\\
0.3	1.03140642318758\\
0.4	1.03140642318758\\
0.5	1.03202124169079\\
0.6	1.03210931312661\\
0.7	1.03213868365553\\
0.8	1.03213868365553\\
0.9	1.03213868365553\\
1	1.03213868365553\\
}--cycle;
\addplot [color=white!55!green, forget plot]
  table[row sep=crcr]{%
1	0.963521590317077\\
0.9	0.963521590317077\\
0.8	0.963521590317077\\
0.7	0.963521590317077\\
0.6	0.963554926784675\\
0.5	0.963654890614917\\
0.4	0.964352728817666\\
0.3	0.964352728817666\\
0.2	0.963821345245808\\
0.1	0.963821345245808\\
0	0.963821345245808\\
};
\addplot [color=white!55!green, forget plot]
  table[row sep=crcr]{%
1	1.03213868365553\\
0.9	1.03213868365553\\
0.8	1.03213868365553\\
0.7	1.03213868365553\\
0.6	1.03210931312661\\
0.5	1.03202124169079\\
0.4	1.03140642318758\\
0.3	1.03140642318758\\
0.2	1.0318745896637\\
0.1	1.0318745896637\\
0	1.0318745896637\\
};
\addplot [color=green]
  table[row sep=crcr]{%
1	0.997830136986301\\
0.9	0.997830136986301\\
0.8	0.997830136986301\\
0.7	0.997830136986301\\
0.6	0.997832119955643\\
0.5	0.997838066152855\\
0.4	0.997879576002623\\
0.3	0.997879576002623\\
0.2	0.997847967454753\\
0.1	0.997847967454753\\
0	0.997847967454753\\
};
\addlegendentry{$\alpha\text{=0.95 (NCD)}$}

\end{axis}
\end{tikzpicture}%
}\hfill
\subfigure[ARI with $\beta$-annealing, $\varepsilon=0.4$\label{fig:results:ARI4}]{
%
%
\begin{tikzpicture}

\begin{axis}[%
width=\figheight,
height=\figwidth,
at={(0.552in,0.48in)},
scale only axis,
xmin=0,
xmax=1,
xlabel style={font=\color{white!15!black}},
xlabel={$\beta$},
ymin=-0.1,
ymax=1.2,
ylabel style={font=\color{white!15!black},at={(0.1,0.5)}},
ylabel={ARI},
axis background/.style={fill=white},
legend style={at={(0.97,0.03)}, anchor=south east, legend cell align=left, align=left, draw=white!15!black}
]

\addplot[area legend, draw=none, fill=blue, fill opacity=0.15, forget plot]
table[row sep=crcr] {%
x	y\\
1	0.509785536332968\\
0.9	0.586835714413143\\
0.8	0.628686657792248\\
0.7	0.667080708039198\\
0.6	0.702133651912282\\
0.5	0.736073969409249\\
0.4	0.743512922767535\\
0.3	0.769085290759585\\
0.2	0.772451202623992\\
0.1	0.770771980889141\\
0	0.754199547125829\\
0	1.12967626336004\\
0.1	1.12812748993234\\
0.2	1.12703006920896\\
0.3	1.12418018415686\\
0.4	1.12612415634142\\
0.5	1.12621211513505\\
0.6	1.12124603513494\\
0.7	1.11731731855344\\
0.8	1.1066122928493\\
0.9	1.08256614935365\\
1	1.02745619297869\\
}--cycle;
\addplot [color=white!55!blue, forget plot]
  table[row sep=crcr]{%
1	0.509785536332968\\
0.9	0.586835714413143\\
0.8	0.628686657792248\\
0.7	0.667080708039198\\
0.6	0.702133651912282\\
0.5	0.736073969409249\\
0.4	0.743512922767535\\
0.3	0.769085290759585\\
0.2	0.772451202623992\\
0.1	0.770771980889141\\
0	0.754199547125829\\
};
\addplot [color=white!55!blue, forget plot]
  table[row sep=crcr]{%
1	1.02745619297869\\
0.9	1.08256614935365\\
0.8	1.1066122928493\\
0.7	1.11731731855344\\
0.6	1.12124603513494\\
0.5	1.12621211513505\\
0.4	1.12612415634142\\
0.3	1.12418018415686\\
0.2	1.12703006920896\\
0.1	1.12812748993234\\
0	1.12967626336004\\
};
\addplot [color=blue]
  table[row sep=crcr]{%
1	0.768620864655828\\
0.9	0.834700931883396\\
0.8	0.867649475320773\\
0.7	0.892199013296321\\
0.6	0.911689843523609\\
0.5	0.931143042272149\\
0.4	0.934818539554479\\
0.3	0.946632737458221\\
0.2	0.949740635916479\\
0.1	0.949449735410739\\
0	0.941937905242932\\
};

\addplot[area legend, draw=none, fill=red, fill opacity=0.15, forget plot]
table[row sep=crcr] {%
x	y\\
1	0.849272493260281\\
0.9	0.850020557015766\\
0.8	0.850184611759248\\
0.7	0.849620878996705\\
0.6	0.848113535745359\\
0.5	0.848204591014771\\
0.4	0.847408844272299\\
0.3	0.84690665866258\\
0.2	0.846596540861124\\
0.1	0.846087053143841\\
0	0.845760704102179\\
0	1.0939220227219\\
0.1	1.09372561893985\\
0.2	1.09341730868643\\
0.3	1.09322945030685\\
0.4	1.09292759368453\\
0.5	1.09245571152916\\
0.6	1.0925081978748\\
0.7	1.09160367651298\\
0.8	1.09094956214804\\
0.9	1.09087492324442\\
1	1.09039865741189\\
}--cycle;
\addplot [color=white!55!red, forget plot]
  table[row sep=crcr]{%
1	0.849272493260281\\
0.9	0.850020557015766\\
0.8	0.850184611759248\\
0.7	0.849620878996705\\
0.6	0.848113535745359\\
0.5	0.848204591014771\\
0.4	0.847408844272299\\
0.3	0.84690665866258\\
0.2	0.846596540861124\\
0.1	0.846087053143841\\
0	0.845760704102179\\
};
\addplot [color=white!55!red, forget plot]
  table[row sep=crcr]{%
1	1.09039865741189\\
0.9	1.09087492324442\\
0.8	1.09094956214804\\
0.7	1.09160367651298\\
0.6	1.0925081978748\\
0.5	1.09245571152916\\
0.4	1.09292759368453\\
0.3	1.09322945030685\\
0.2	1.09341730868643\\
0.1	1.09372561893985\\
0	1.0939220227219\\
};
\addplot [color=red]
  table[row sep=crcr]{%
1	0.969835575336084\\
0.9	0.970447740130091\\
0.8	0.970567086953642\\
0.7	0.970612277754843\\
0.6	0.970310866810079\\
0.5	0.970330151271966\\
0.4	0.970168218978414\\
0.3	0.970068054484714\\
0.2	0.970006924773775\\
0.1	0.969906336041846\\
0	0.969841363412041\\
};

\addplot[area legend, draw=none, fill=green, fill opacity=0.15, forget plot]
table[row sep=crcr] {%
x	y\\
1	1\\
0.9	1\\
0.8	1\\
0.7	1\\
0.6	1\\
0.5	1\\
0.4	1\\
0.3	1\\
0.2	1\\
0.1	1\\
0	1\\
0	1\\
0.1	1\\
0.2	1\\
0.3	1\\
0.4	1\\
0.5	1\\
0.6	1\\
0.7	1\\
0.8	1\\
0.9	1\\
1	1\\
}--cycle;
\addplot [color=white!55!green, forget plot]
  table[row sep=crcr]{%
1	1\\
0.9	1\\
0.8	1\\
0.7	1\\
0.6	1\\
0.5	1\\
0.4	1\\
0.3	1\\
0.2	1\\
0.1	1\\
0	1\\
};
\addplot [color=white!55!green, forget plot]
  table[row sep=crcr]{%
1	1\\
0.9	1\\
0.8	1\\
0.7	1\\
0.6	1\\
0.5	1\\
0.4	1\\
0.3	1\\
0.2	1\\
0.1	1\\
0	1\\
};
\addplot [color=green]
  table[row sep=crcr]{%
1	1\\
0.9	1\\
0.8	1\\
0.7	1\\
0.6	1\\
0.5	1\\
0.4	1\\
0.3	1\\
0.2	1\\
0.1	1\\
0	1\\
};

\end{axis}
\end{tikzpicture}%
}\hfill
\subfigure[ARI with $\beta$-annealing, $\varepsilon=0.8$\label{fig:results:ARI8}]{
%
%
\begin{tikzpicture}

\begin{axis}[%
width=\figheight,
height=\figwidth,
at={(0.603in,0.48in)},
scale only axis,
xmin=0,
xmax=1,
xlabel style={font=\color{white!15!black}},
xlabel={$\beta$},
ymin=-0.1,
ymax=1.2,
ylabel style={font=\color{white!15!black},at={(0.1,0.5)}},
ylabel={ARI},
axis background/.style={fill=white},
]

\addplot[area legend, draw=none, fill=blue, fill opacity=0.15, forget plot]
table[row sep=crcr] {%
x	y\\
1	0.0371899431591959\\
0.9	0.0296375105251184\\
0.8	0.021137462515768\\
0.7	0.0556120920688029\\
0.6	0.121615013775018\\
0.5	0.142590393015519\\
0.4	0.143804884413238\\
0.3	0.130669979273094\\
0.2	0.1092005707366\\
0.1	0.0560137817604774\\
0	-0.0424882072545114\\
0	0.450302861162674\\
0.1	0.67748825627721\\
0.2	0.799926468660685\\
0.3	0.856890709294356\\
0.4	0.88591789599112\\
0.5	0.873145556004866\\
0.6	0.823040356652685\\
0.7	0.688031801529895\\
0.8	0.536739578657266\\
0.9	0.445135317177927\\
1	0.399542083744065\\
}--cycle;
\addplot [color=white!55!blue, forget plot]
  table[row sep=crcr]{%
1	0.0371899431591959\\
0.9	0.0296375105251184\\
0.8	0.021137462515768\\
0.7	0.0556120920688029\\
0.6	0.121615013775018\\
0.5	0.142590393015519\\
0.4	0.143804884413238\\
0.3	0.130669979273094\\
0.2	0.1092005707366\\
0.1	0.0560137817604774\\
0	-0.0424882072545114\\
};
\addplot [color=white!55!blue, forget plot]
  table[row sep=crcr]{%
1	0.399542083744065\\
0.9	0.445135317177927\\
0.8	0.536739578657266\\
0.7	0.688031801529895\\
0.6	0.823040356652685\\
0.5	0.873145556004866\\
0.4	0.88591789599112\\
0.3	0.856890709294356\\
0.2	0.799926468660685\\
0.1	0.67748825627721\\
0	0.450302861162674\\
};
\addplot [color=blue]
  table[row sep=crcr]{%
1	0.21836601345163\\
0.9	0.237386413851522\\
0.8	0.278938520586517\\
0.7	0.371821946799349\\
0.6	0.472327685213852\\
0.5	0.507867974510192\\
0.4	0.514861390202179\\
0.3	0.493780344283725\\
0.2	0.454563519698642\\
0.1	0.366751019018844\\
0	0.203907326954081\\
};

\addplot[area legend, draw=none, fill=red, fill opacity=0.15, forget plot]
table[row sep=crcr] {%
x	y\\
1	0.542945385671188\\
0.9	0.60318996272219\\
0.8	0.6590133625715\\
0.7	0.717913075144273\\
0.6	0.772375684392222\\
0.5	0.788696714526617\\
0.4	0.793615444825542\\
0.3	0.785956890482047\\
0.2	0.75890740228559\\
0.1	0.704040333772763\\
0	0.506663787395868\\
0	1.03195792929197\\
0.1	1.06001852200283\\
0.2	1.07595286994393\\
0.3	1.08905491845017\\
0.4	1.09257857985859\\
0.5	1.09438856150656\\
0.6	1.09070308054128\\
0.7	1.08891572155777\\
0.8	1.07449300270977\\
0.9	1.03778147970345\\
1	0.988444739783586\\
}--cycle;
\addplot [color=white!55!red, forget plot]
  table[row sep=crcr]{%
1	0.542945385671188\\
0.9	0.60318996272219\\
0.8	0.6590133625715\\
0.7	0.717913075144273\\
0.6	0.772375684392222\\
0.5	0.788696714526617\\
0.4	0.793615444825542\\
0.3	0.785956890482047\\
0.2	0.75890740228559\\
0.1	0.704040333772763\\
0	0.506663787395868\\
};
\addplot [color=white!55!red, forget plot]
  table[row sep=crcr]{%
1	0.988444739783586\\
0.9	1.03778147970345\\
0.8	1.07449300270977\\
0.7	1.08891572155777\\
0.6	1.09070308054128\\
0.5	1.09438856150656\\
0.4	1.09257857985859\\
0.3	1.08905491845017\\
0.2	1.07595286994393\\
0.1	1.06001852200283\\
0	1.03195792929197\\
};
\addplot [color=red]
  table[row sep=crcr]{%
1	0.765695062727387\\
0.9	0.820485721212822\\
0.8	0.866753182640633\\
0.7	0.903414398351021\\
0.6	0.93153938246675\\
0.5	0.94154263801659\\
0.4	0.943097012342065\\
0.3	0.93750590446611\\
0.2	0.91743013611476\\
0.1	0.882029427887798\\
0	0.769310858343919\\
};

\addplot[area legend, draw=none, fill=green, fill opacity=0.15, forget plot]
table[row sep=crcr] {%
x	y\\
1	0.964716935413275\\
0.9	0.964716935413275\\
0.8	0.964716935413275\\
0.7	0.964716935413275\\
0.6	0.964352728817666\\
0.5	0.963821345245808\\
0.4	0.964352728817666\\
0.3	0.964716935413275\\
0.2	0.9651461905661\\
0.1	0.9651461905661\\
0	0.9651461905661\\
0	1.03070735718258\\
0.1	1.03070735718258\\
0.2	1.03070735718258\\
0.3	1.03108554514867\\
0.4	1.03140642318758\\
0.5	1.0318745896637\\
0.6	1.03140642318758\\
0.7	1.03108554514867\\
0.8	1.03108554514867\\
0.9	1.03108554514867\\
1	1.03108554514867\\
}--cycle;
\addplot [color=white!55!green, forget plot]
  table[row sep=crcr]{%
1	0.964716935413275\\
0.9	0.964716935413275\\
0.8	0.964716935413275\\
0.7	0.964716935413275\\
0.6	0.964352728817666\\
0.5	0.963821345245808\\
0.4	0.964352728817666\\
0.3	0.964716935413275\\
0.2	0.9651461905661\\
0.1	0.9651461905661\\
0	0.9651461905661\\
};
\addplot [color=white!55!green, forget plot]
  table[row sep=crcr]{%
1	1.03108554514867\\
0.9	1.03108554514867\\
0.8	1.03108554514867\\
0.7	1.03108554514867\\
0.6	1.03140642318758\\
0.5	1.0318745896637\\
0.4	1.03140642318758\\
0.3	1.03108554514867\\
0.2	1.03070735718258\\
0.1	1.03070735718258\\
0	1.03070735718258\\
};
\addplot [color=green]
  table[row sep=crcr]{%
1	0.997901240280973\\
0.9	0.997901240280973\\
0.8	0.997901240280973\\
0.7	0.997901240280973\\
0.6	0.997879576002623\\
0.5	0.997847967454753\\
0.4	0.997879576002623\\
0.3	0.997901240280973\\
0.2	0.997926773874341\\
0.1	0.997926773874341\\
0	0.997926773874341\\
};

\end{axis}
\end{tikzpicture}%
}
\caption{(a)-(c): Colorplots of the transition probability matrices $\Pvec$ for different values of $\alpha$ and $\varepsilon$. For large $\alpha$ the block diagonal structure becomes more dominant. (d): A random permutation of the rows and columns hides the block structure. (e)-(j): Curves showing the cost function $\cost_\beta$ and the adjusted Rand index (ARI) for different settings. Mean values (solid lines) are shown together with the standard deviation (shaded areas).}
\label{fig:results}
\vskip -0.2in
\end{figure*}

\newcommand{\ri}{\mathsf{ARI}}
\section{Experiments and Results}\label{sec:experiments}
\subsection{A Non-Reversible Markov Chain}\label{sec:experiments:reversibility}
The last property of Lemma~\ref{lem:costbeta_properties} cannot be generalized to non-reversible Markov chains. Specifically, as the proof of Lemma~\ref{lem:costbeta_properties} shows, $\cost_\beta$ is non-decreasing in $\beta$ iff $\cost_P \ge 2\cost_L$. Since one can find also non-reversible Markov chains for which this holds, reversibility is sufficient but not necessary for $\cost_\beta$ to be non-decreasing in $\beta$. We next consider a non-reversible Markov chain $\Xvec\sim\Mar{\{1,2,3\},\Pvec}$ with
\begin{equation}
 \Pvec = \left[ \begin{array}{ccc}
                 0.4 & 0.3 & 0.3 \\ 0.25 & 0.3 & 0.45 \\ 0.15 & 0.425 & 0.425
                \end{array}\right]
\end{equation}
and let $g$ be such that $g(1)=1$ and $g(2)=g(3)=2$. Then, $\cost_L = 0.0086$ and $\cost_P = 0.0135$, i.e., $\cost_P < 2\cost_L$. In this case, $\cost_\beta$ is decreasing with increasing $\beta$.

\subsection{Quasi-Lumpable and Nearly Completely Decomposable Markov Chains}\label{sec:experiments:QL}
Suppose we have a partition $\{\dom{X}_i\}$, $i=1,\dots,M$, of $\dom{X}$ with $\card{\dom{X}_i}=N_i$. Then for any $\Amat'$ and $\Pvec'_{ij}$ that are  $M \times M$ and $N_i\times N_j$ row stochastic matrices, respectively, define $\Amat=[a_{ij}]=(1-\alpha) \Amat' + \alpha \eye$, $\alpha\in[0,1]$, and let
\begin{equation}\label{eq:blockstochastic}
 \Pvec' = \left[
 \begin{array}{cccc}
  a_{11} \Pvec'_{11} & a_{12}\Pvec'_{12}  &\cdots & a_{1M}\Pvec'_{1M}\\
  a_{21} \Pvec'_{21} & a_{22}\Pvec'_{22}  &\cdots & a_{2M}\Pvec'_{2M}\\
  \vdots & \vdots & \ddots & \vdots\\
  a_{M1} \Pvec'_{M1} & a_{M2}\Pvec'_{22}  &\cdots & a_{MM}\Pvec'_{MM}
\end{array}
 \right].
\end{equation}
Let further $\Xvec\sim\Mar{\dom{X},\Pvec'}$. If $g$ induces the partition $\{\dom{X}_i\}$, then it can be shown that $\Yvec$ is Markov with transition probability matrix $\Amat$, i.e., $\Marx{\Yvec}\equiv\Yvec$, and $\cost_0(\Xvec,\Marx{\Yvec})=0$. The Markov chain $\Xvec$ is \emph{lumpable} w.r.t.\ the partition $g$. The matrix $\Pvec'$ is block stochastic and the parameter $\alpha$ specifies how dominant the diagonal blocks are. Specifically, if $\alpha=1$, then $\Pvec'$ is block diagonal and we call $\Xvec$ \emph{completely decomposable}. Such a Markov chain is not irreducible. We hence look at Markov chains $\Xvec\sim\Mar{\dom{X},\Pvec}$ with 
\begin{equation}
 \Pvec=(1-\varepsilon)\Pvec'+ \varepsilon \Emat
\end{equation}
where $\varepsilon\in[0,1]$ and where $\Emat$ (which can be interpreted as noise) is row stochastic and irreducible. For small values of $\varepsilon$ we call $\Xvec$ nearly completely decomposable (NCD) if $\alpha$ is close to one, otherwise we call it quasi-lumpable.

We now perform experiments with these types of Markov chains. We set $M=3$, $N_1=N_2=25$, and $N_3=50$, and chose the parameters from $\alpha\in\{0, 0.5, 0.95\}$ and $\varepsilon\in\{0, 0.4, 0.8\}$. For each pair $(\alpha,\varepsilon)$, we generated 250 random matrices $\Amat'$ and $\Pvec'_{ij}$. A selection of the corresponding matrices $\Pvec$ is shown in Fig.~\ref{fig:results}.

Note that in practice the states of even a completely decomposable Markov chain $\Xvec$ are rarely ordered such that the transition probability matrix is block diagonal. Rather, the state labeling must be assumed to be random. In this case, $\Pvec$ is obtained by a random permutation of the rows and columns of a block diagonal matrix (see Fig.~\ref{fig:results:perm}), which prevents the optimal aggregation function being ``read off'' simply by looking at $\Pvec$. That $\Pvec$ has a block structure in our case does not affect the performance of our algorithms, since they 1) are unaware of this structure and 2) are initialized randomly.

\begin{table*}[t]
 \caption{Aggregating a letter bi-gram model. The partitions are shown together with the ARI $\ri$ w.r.t.\ the reference partition (first row) for $\card{\mathcal{Y}} =4$} 
 \label{tab:bigram1}
 \centering
 \begin{tabular}{c|c|l}
  $\beta$ Value & $\ri$ & Partitions, shown for $\card{\mathcal{Y}} =4$\\
    \hline 
    Ref. & -- & \texttt{\{\textvisiblespace\},\{!"\$'(),-.:;?[]\},\{aeiou\},\{0123456789\},\{AEIOU\},\{BCDFGHJKLMNPQRSTVWYZ\},\{bcdfghjklmnpqrstvwxyz\}}\\
    \hline 
    $\beta=1$ & $0.43$& \texttt{\{\textvisiblespace!'),-.0:;?]\},\{aeioy\},\{"\$(123456789ABCDEFGHIJKLMNOPQRSTUVWY[h\},\{Zbcdfgjklmnpqrstuvwxz\}}\\
  \hline 
    $\beta={0.8}$ & $0.46$& \texttt{\{\textvisiblespace!'),-.:;?Z]\},\{aeiouy\},\{"\$(0123456789ABCDEFGHIJKLMNOPQRSTUVWY[h\},\{bcdfgjklmnpqrstvwxz\}}\\
  \hline
      $\beta={0.5}$ & $0.35$& \texttt{\{\textvisiblespace!3?Z\},\{'2456789AOUaeiou\},\{"\$(-01BCDEFGHIJKLMNPQRSTVWY[bhjqw\},\{),.:;]cdfgklmnprstvxyz\}}\\
  \hline
  $\beta={0}$ & $0.12$& \texttt{\{\textvisiblespace-2CEFMPSTcfgopst\},\{"'456789AOUZaeiu\},\{!\$1?BDGHJLNQRVW[bhjklmqrvwz\},\{(),.03:;IKY]dnxy\}}\\
 \end{tabular}
\end{table*}

\begin{table*}[t]
	\caption{Aggregating a letter bi-gram model. The partitions are shown together with the ARI $\ri$ w.r.t.\ the reference partition (first row) for $\card{\mathcal{Y}} \in \{2,7\}$}
	\label{tab:bigram2}
	\centering
	\begin{tabular}{c|c|l}
	 $\beta$ Value & $\ri$ & Partitions, shown for $\card{\mathcal{Y}} \in \{2,7\}$\\
		\hline 
		Ref. & -- & \texttt{\{\textvisiblespace\},\{!"\$'(),-.:;?[]\},\{aeiou\},\{0123456789\},\{AEIOU\},\{BCDFGHJKLMNPQRSTVWYZ\},\{bcdfghjklmnpqrstvwxyz\}}\\
		\hline 
		$\beta=1$ & $0.2$& \texttt{\{\textvisiblespace!"'),-.01235689:;?KU]aehioy\},\{\$(47ABCDEFGHIJLMNOPQRSTVWYZ[bcdfgjklmnpqrstuvwxz\}} \\
		  & $0.34$ & \texttt{\{\textvisiblespace!'),-.:;?]\},\{aeioy\},\{"\$(0123456789ABCDEFGHIJLMNOPQRSTVWY[\},\{bcfjmpqstw\},\{dgx\},\{KUh\},\{Zklnruvz\}}\\
		\hline 
		$\beta={0.8}$ & $0.24$& \texttt{\{\textvisiblespace!"'),-.01235689:;?EU]aehiouy\},\{\$(47ABCDFGHIJKLMNOPQRSTVWYZ[bcdfgjklmnpqrstvwxz\}}\\
		  & $0.35$ & \texttt{\{\textvisiblespace!'),-.:;?]\},\{aeioy\},\{\$"(0123456789ABCDEFGHIJLMNOPQRSTUVWYZ[\},\{Kh\},\{bcfjkmpqstw\},\{dg\},\{lnruvxz\}}\\
		\hline
		$\beta={0.5}$ & $0.15$& \texttt{\{\textvisiblespace!'-12368?EOUZaeiou\},\{"\$(),.04579:;ABCDFGHIJKLMNPQRSTVWY[]bcdfghjklmnpqrstvwxyz\}}\\
		  & $0.31$  & \texttt{\{\textvisiblespace'\},\{!),.:;?]dy\},\{aeiou\},\{"\$(-0123589ACEIMOPRSTUWZ\},\{BDFGHJKLNQVYhj\},\{7[bcfgkmpqstw\},\{46lnrvxz\}}  \\
		\hline
		$\beta={0}$ & $0.01$& \texttt{\{\textvisiblespace!\$(-0124578?ABCFHLMNOPRSTUVWaceglnostuwxz\},\{"'),.369:;DEGIJKQYZ[]bdfhijkmpqrvy\}}\\
		  & $0.02$ & \texttt{\{\textvisiblespace4689ao\},\{\$'AKOiux\},\{!?HVZhjkmvz\},\{"(-25CEFLMNRUWY[egnprs\},\{37BPQbl\},\{1:;STctw\},\{),.0DGIJ]dfy\}}  \\
	\end{tabular}
\end{table*}

%
%

We applied our aggregation algorithm both with and without the annealing procedure for $\beta\in\{0,0.1,\dots,0.9,1\}$ with the goal of retrieving the partition $\{\dom{X}_i\}$. We measure the success, i.e., the degree to which the function $g$ obtained from the algorithm agrees with the partition $\{\dom{X}_i\}$, using the adjusted Rand index (ARI). An ARI of one indicates that the two partitions are equivalent. Note that we always assume that the number $M$ of sets in the partition $\{\dom{X}_i\}$ is known.

The results are shown in Fig.~\ref{fig:results}. Specifically, Fig.~\ref{fig:results:cost} shows that the cost for the aggregation found by our algorithm with $\beta$-annealing decreases monotonically with decreasing $\beta$: We obtain a partition for a given value of $\beta$. This partition has, by assuming $\cost_P\ge 2\cost_L$ (cf.~Section~\ref{sec:experiments:reversibility}), an even lower cost for a smaller value of $\beta$. Further optimization for this smaller value of $\beta$ further reduces the cost, leading to the depicted phenomenon. In contrast, the sequential Algorithm~\ref{alg:sGITMA} without the annealing procedure fails for values of $\beta$ less than $0.5$. This is apparent both in the cost in Fig.~\ref{fig:results:costseq} (which has a sharp jump around $\beta=0.5$) and in the ARI in Fig.~\ref{fig:results:ARIseq} (which drops to zero). Apparently, the algorithm gets stuck in a bad local optimum.

Figs.~\ref{fig:results:ARI0} to~\ref{fig:results:ARI8} show the ARI of the aggregations obtained by our algorithm with $\beta$-annealing. First of all, it can be seen that performance improves with increasing $\alpha$, since the dominant block structure makes discovering the correct partition more easy. Moreover, it can be seen that for $\alpha=0$ the optimum $\beta$ lies at smaller values, typically smaller than 0.5. The position of this optimum increases with increasing noise: While in the noiseless case the correct partition is typically obtained for $\beta$ close to zero, in the highly noisy case of $\varepsilon=0.8$ we require $\beta\approx 0.4$ to achieve good results. The reason may be that the higher noise leads to more partitions being quasi-lumpable by leading to an i.i.d.\  $\Yvec$, hence for small values of $\beta$ one may get drawn into these ``false solutions'' more easily. In contrast, for NCD Markov chains (i.e., for $\alpha=0.95$) sometimes noise helps in discovering the correct partition. Comparing Figs.~\ref{fig:results:ARI0} and~\ref{fig:results:ARI4}, one can see that a noise of $\varepsilon=0.4$ allows us to perfectly discover the partition. We believe that a small amount of noise helps in escaping bad local minima.

The fact that the $\beta$ for which the highest ARI is achieved not necessarily falls together with the values 0, 0.5, or 1 indicates that our generalized aggregation framework has the potential to strictly outperform aggregation cost functions and algorithms that have been previously proposed (cf.~Section~\ref{sec:related}).

\begin{figure*}[t]
\subfigure[$\Pvec$, $k=15$]{\includegraphics[width=0.20\textwidth, trim=1.5cm 0cm 1.5cm 0.5cm, clip]{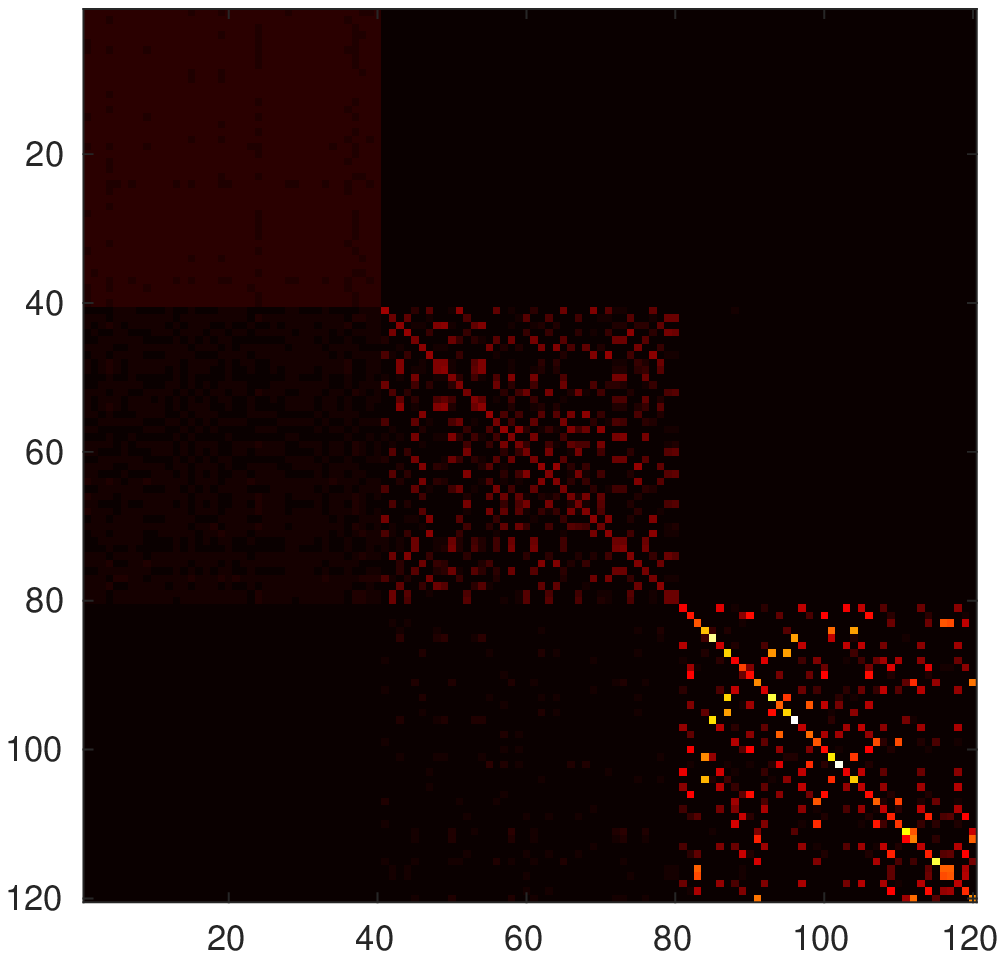}
}\hfill
\subfigure[$\beta\in\{0.2,0.5,0.8\}$, $k=15$]{
\begin{pspicture}[showgrid=false](-1,-1.5)(2,2)
\psset{yunit=0.1,xunit=0.1}
\readdata{\data}{3C_15_5_1.dat}
\rput(4,4){\listplot[plotstyle=dots,linecolor=red]{\data}}
\readdata{\data}{3C_15_5_2.dat}
\rput(4,4){\listplot[plotstyle=dots,linecolor=orange!50!yellow]{\data}}
\readdata{\data}{3C_15_5_3.dat}
\rput(4,4){\listplot[plotstyle=dots,linecolor=black]{\data}}
\end{pspicture}
}\hfill
\subfigure[$\Pvec$, $k=120$]{\includegraphics[width=0.20\textwidth, trim=1.5cm 0cm 1.5cm 0.5cm, clip]{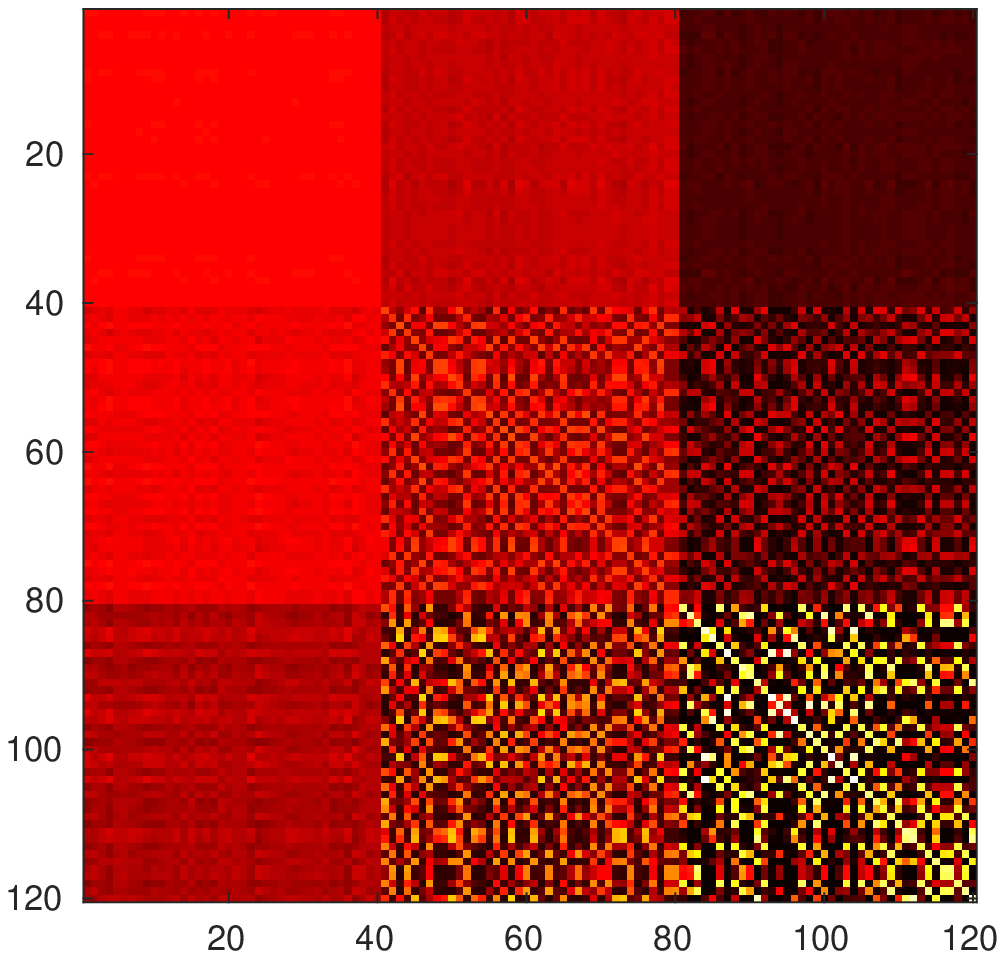}
}\hfill
\subfigure[$\beta=0.2$, $k=120$]{
\begin{pspicture}[showgrid=false](-1,-1.5)(2,2)
\psset{yunit=0.1,xunit=0.1}
\readdata{\data}{3C_120_2_1.dat}
\rput(4,4){\listplot[plotstyle=dots,linecolor=red]{\data}}
\readdata{\data}{3C_120_2_2.dat}
\rput(4,4){\listplot[plotstyle=dots,linecolor=orange!50!yellow]{\data}}
\readdata{\data}{3C_120_2_3.dat}
\rput(4,4){\listplot[plotstyle=dots,linecolor=black]{\data}}
\end{pspicture}
}\hfill
\subfigure[$\beta=0.8$, $k=120$]{
\begin{pspicture}[showgrid=false](-1,-1.5)(2,2)
\psset{yunit=0.1,xunit=0.1}
\readdata{\data}{3C_120_8_1.dat}
\rput(4,4){\listplot[plotstyle=dots,linecolor=red]{\data}}
\readdata{\data}{3C_120_8_2.dat}
\rput(4,4){\listplot[plotstyle=dots,linecolor=orange!50!yellow]{\data}}
\readdata{\data}{3C_120_8_3.dat}
\rput(4,4){\listplot[plotstyle=dots,linecolor=black]{\data}}
\end{pspicture}
}\\
\subfigure[$\Pvec$, $k=15$]{\includegraphics[width=0.20\textwidth, trim=1.5cm 0cm 1.5cm 0.5cm, clip]{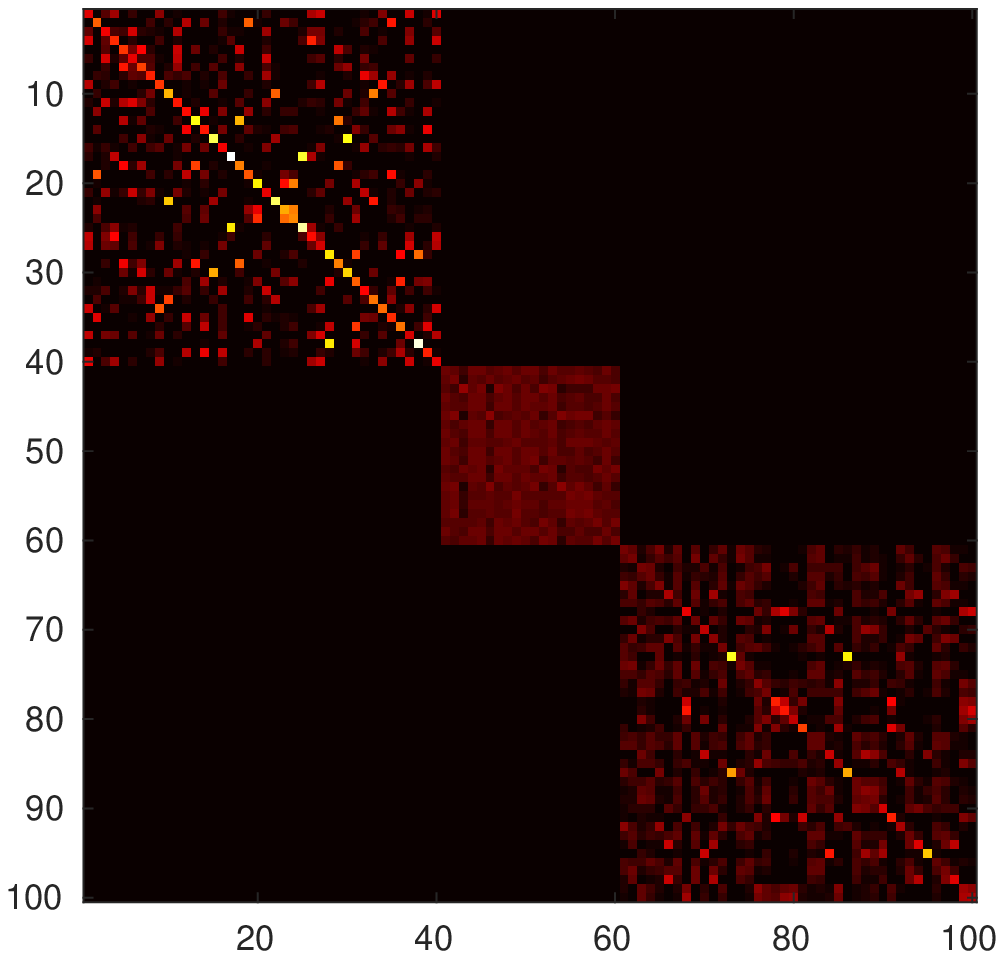}
}\hfill
\subfigure[$\Pvec$, $k=100$]{\includegraphics[width=0.20\textwidth, trim=1.5cm 0cm 1.5cm 0.5cm, clip]{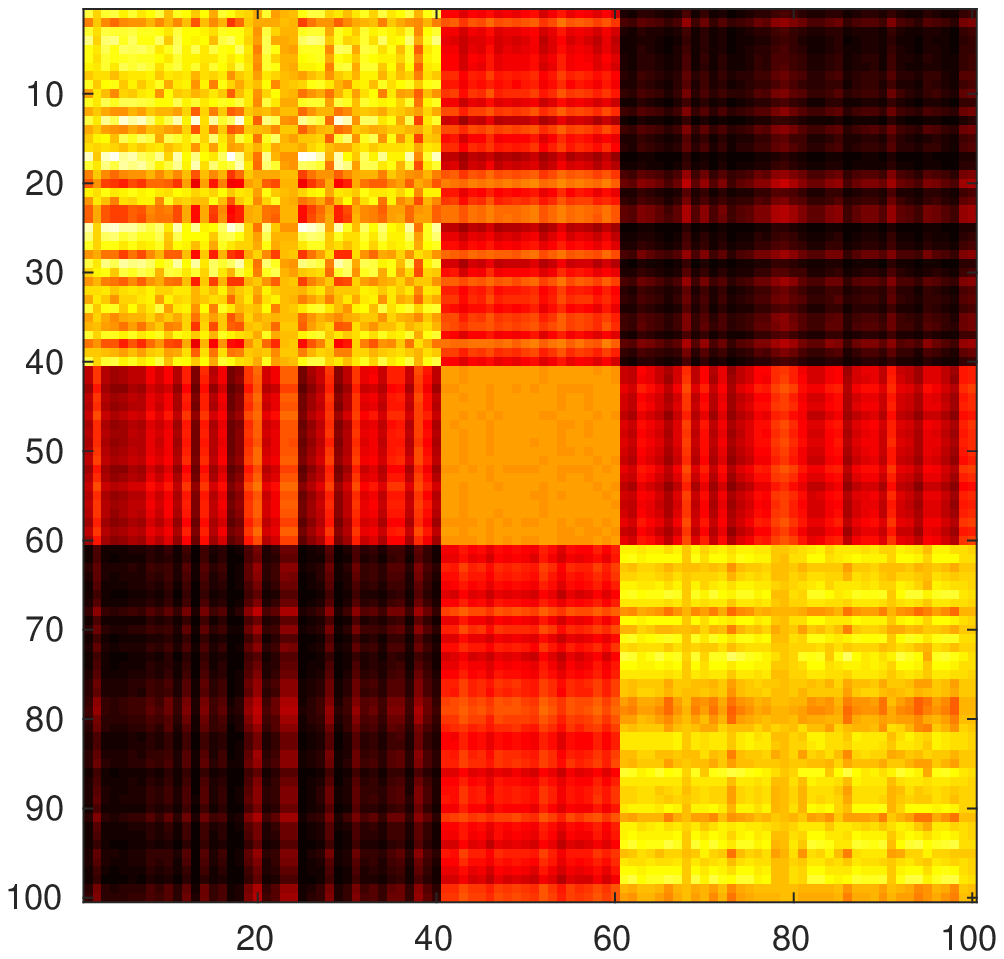}
}\hfill
\subfigure[$\beta\in\{0.2,0.5,0.8\}$, $k\in\{15,100\}$]{
\begin{pspicture}[showgrid=false](-2,-1.5)(4,2)
\psset{yunit=0.2,xunit=0.2}
\readdata{\data}{clouds_15_2_1.dat}
\rput(4,4){\listplot[plotstyle=dots,linecolor=red]{\data}}
\readdata{\data}{clouds_15_2_2.dat}
\rput(4,4){\listplot[plotstyle=dots,linecolor=orange!50!yellow]{\data}}
\readdata{\data}{clouds_15_2_3.dat}
\rput(4,4){\listplot[plotstyle=dots,linecolor=black]{\data}}
\end{pspicture}
}\hfill
\caption{Clustering three circles (first row) and three linearly separable clusters (second row). For $k=15$, the transition probability matrices (shown in (a) and (f)) are nearly completely decomposable. The result for the three circles depends strongly on a careful setting of the parameters $\beta$ and $k$ ((b), (d), and (e)), while the three linearly separable clusters were separated correctly for all parameter choices (h).}
\label{fig:pairwiseclustering}
\vskip -0.2in
\end{figure*} 

\subsection{An Example from Natural Language Processing}
We took the letter bi-gram model from~\cite{GeigerWu_HigherOrder}, which was obtained by analyzing the co-occurrence of letters in F. Scott Fitzgerald's book ``The Great Gatsby''. The text was modified by removing chapter headings, line breaks, underscores, and by replacing \texttt{\'e} by \texttt{e}. With the remaining symbols, we obtained a Markov chain with an alphabet size of $N=76$ (upper and lower case letters, numbers, punctuation, etc.).

We applied Algorithm~\ref{alg:Ann} for $\card{\mathcal{Y}}\in\{2,\dots,7\}$ and $\beta\in\{0,0.1,\dots,0.9,1\}$. To get consistent results, we restarted the algorithm 20 times for $\beta=1$ and chose the aggregation $g$ that minimized $\cost_1(\Xvec,g)$; we used this aggregation $g$ as an initialization for the $\beta$-annealing procedure.

Looking at the results for $\card{\mathcal{Y}} =4$ in Table~\ref{tab:bigram1}, one can observe that the results for $\beta =0.8$ appear to be most meaningful when compared to other values of $\beta$ such as $\beta=1$ (information bottleneck), $\beta=0.5$ (as proposed in~\cite{Meyn_MarkovAggregation}), and $\beta=0$ (as proposed in~\cite{GeigerEtAl_OptimalMarkovAggregation}). Specifically, for $\beta=0$ it can be seen that not even the annealing procedure was able to achieve meaningful results. This conclusion is supported by calculating the ARI of these aggregations for a plausible reference aggregation of the alphabet into upper case vowels, upper case consonants, lower case vowels, lower case consonants, numbers, punctuation, and the blank space as shown in the first row of the Table~\ref{tab:bigram1}. The absolute ARI values are not a good performance indicator in this case since we are comparing to a reference partition with seven sets whereas $\card{\mathcal{Y}} =4$.

In Table~\ref{tab:bigram2} the same experiment is repeated for $\card{\mathcal{Y}} \in \{2,7\}$. We again observe that $\beta=0.8$ leads to the most meaningful results which is also supported by ARI values. 

\subsection{Clustering via Markov Aggregation}
Data points are often described only by pairwise similarity values, and these similarity values can be used to construct the transition probability matrix of a Markov chain. Then, with this probabilistic interpretation, our information-theoretic cost functions for Markov aggregation can be used for clustering. This approach has been taken by~\cite{Alush_PairwiseClustering,Tishby_MarkovRelaxation}.

We considered two different data sets: three linearly separable clusters and three concentric circles, as shown in Fig~\ref{fig:pairwiseclustering}. The three linearly separable clusters were obtained by placing 40, 20, and 40 points, drawn from circularly symmetric Gaussian distributions with standard deviations 2.5, 0.5, and 1.5 at horizontal coordinates -10, 0, and 10, respectively. The three concentric circles were obtained by placing 40 points each at uniformly random angles at radii $\{0.1, 7, 15\}$, and by adding to each data point spherical Gaussian noise with a standard deviation of $0.3$. In both cases, we computed the transition probability matrix $\Pvec$ according to
\begin{equation}\label{eq:pairwise}
 P_{i\to j} \propto \e{-\frac{\Vert x_i-x_j\Vert^2_2}{\sigma_k} }
\end{equation}
where $x_i$ and $x_j$ are the coordinates of the $i$-th and $j$-th data point, $\Vert\cdot\Vert_2^2$ is the squared Euclidean distance, and where $\sigma_k$ is a scale parameter. We set $\sigma_k$ to the average squared Euclidean distance between a data point and its $k$ nearest neighbors (and averaged this quantity over all data points). We set $k$ either to 15 or to the total number of data points.

We applied our Algorithm~\ref{alg:Ann} with the annealing procedure for $\beta$. As in the previous experiment, we restarted the algorithm 50 times for $\beta=1$ and chose the aggregation $g$ that minimized $\cost_1(\Xvec,g)$; we used this aggregation $g$ as an initialization for the $\beta$-annealing procedure.

The results are shown in Fig.~\ref{fig:pairwiseclustering}, together with a colorplot of the respective transition probability matrices. It can be seen that the three linearly separable clusters were detected correctly for all chosen parameter values. This is not surprising for $k=15$, since in this case the resulting Markov chain is nearly completely decomposable. Interestingly, though, the same results were observed for $k=100$ for which $\Pvec$ is structured, but not block diagonal. One may claim that these results are due to Algorithm~\ref{alg:Ann} getting stuck in a local optimum for $\beta=1$ which accidentally coincides with the correct clustering, and that optimizing our cost function for values of $\beta$ larger than 0.5 but smaller than 1 may fail. Since we reproduced these results by using Algorithm~\ref{alg:sGITMA} (with 50 restarts to escape bad local optima) for values of $\beta$ greater than 0.5, this claim can be refuted.

For the three concentric circles, things look different. We correctly identified the clusters only for a nearly completely decomposable $\Pvec$, i.e., for a careful setting of $k$ (and we were able to reproduce these results for $\beta$ greater than 0.5 using Algorithm~\ref{alg:sGITMA}). For $k=120$, i.e., equal to the number of data points, the three circles were not identified correctly. 

Since we have reason to believe that the optimal $k$ depends strongly on the data set, we are hesitant to recommend this approach to cluster data points that are not linearly separable (in which case a simpler method such as k-means would suffice). Our preliminary analysis of~\cite{Alush_PairwiseClustering} suggests that their approach (in which $\Xvec$ is a random walk on the $k$-nearest neighbor graph of the data set and in which the authors chose $\beta=0.5$) suffers from similar problems. Finally, the authors of~\cite{Tishby_MarkovRelaxation} suggest to let $\Xvec$ ``relax'' to some metastable point, i.e., take an $r$-th power of $\Pvec$ such that $\Pvec^r$ is approximately a projection; their approach is equivalent to ours for $\beta=1$, with $\Pvec$ replaced by $\Pvec^r$. Nevertheless, also this approach requires setting $r$ and $k$ for~\eqref{eq:pairwise}. Whether this relaxation to metastability can be successfully combined with our generalized cost function for Markov aggregation will be deferred to future investigations.

\section{Proofs}

\subsection{Proof of Proposition~\ref{prop:bisimulation}}\label{proof:bisimulation}
 Consider the relation $\epsrel=\{(g(x),x){:}\ x\in\dom{X}\}$. It can be shown that
 \begin{equation}
  \forall T\subseteq \dom{X}\cup\dom{Y}{:}\ \epsrel(T) = \preim{T\cap \dom{Y}}\subseteq\dom{X}.
 \end{equation}
 We thus need to show that, for all $x$ and all $B\subseteq\dom{Y}$,
 \begin{equation}\label{eq:condBisimilarityConcrete}
  \sum_{x'\in \preim{B}}P_{x\to x'} \ge \sum_{y\in B} Q_{g(x)\to y} - \varepsilon.
 \end{equation}
 Now let $\Rvec=[R_{x\to y}]=\Pvec\Wvec$, i.e., we have
 \begin{equation}\label{eq:defR}
  R_{x\to y} = \sum_{x'\in\preim{y}} P_{x\to x'}.
 \end{equation}
 One can show along the lines of~\cite[(65)--(68)]{GeigerEtAl_OptimalMarkovAggregation} that
 \begin{equation}
  \costl = \sum_{x\in\dom{X}} \mu_x \underbrace{\sum_{y\in\dom{Y}} R_{x\to y} \log \frac{R_{x\to y}}{Q_{g(x)\to y}}}_{=:\kld{R_{x\to \cdot}}{Q_{g(x)\to \cdot}}}
 \end{equation}
 from which we get that, for every $x$,
 \begin{equation}
  \kld{R_{x\to \cdot}}{Q_{g(x)\to \cdot}} \le \frac{\costl}{\min_x \mu_x}.
 \end{equation}
 With Pinsker's inequality~\cite[Lemma~12.6.1]{Cover_Information1} and~\cite[(12.137)]{Cover_Information1} we thus get that, for every $x$ and every $B\subseteq\dom{Y}$,
 \begin{equation}
  \left|\sum_{y\in B} R_{x\to y}-Q_{g(x)\to y} \right| \le \sqrt{\frac{\ln(2) \costl}{2\min_x \mu_x}}.
 \end{equation}
 Combining this with~\eqref{eq:defR} thus shows that~\eqref{eq:condBisimilarityConcrete} holds for 
 \begin{equation}
  \varepsilon = \sqrt{\frac{\ln(2) \costl}{2\min_x \mu_x}}.
 \end{equation}
 This completes the proof.

\subsection{Proof of Lemma~\ref{lem:deltabeta_nondecreasing}}\label{proof:deltabeta_nondecreasing}
 We show that the derivative of $\delta_\beta(\Xvec,\Wvec)$ w.r.t.\ $\beta$ is positive. Indeed,
 \begin{align}
  &\frac{\diff}{\diff\beta}\delta_\beta(\Xvec,\Wvec) \notag\\
  &= \redrate{\Xvec}-\redrate{\Yvec} - \kldr{\Yvec}{\Marx{\Yvec}}\\
  &= \mutinf{X_1;X_2}-\ent{Y}+\entrate{\Yvec} - \ent{Y_2|Y_1} + \entrate{\Yvec}.
 \end{align}
 The entropy rate of the reversed process equals the entropy rate of the original process, i.e.,
 \begin{equation}
  \entrate{\Yvec} = \limn \ent{Y_n|Y_1^{n-1}} = \limn \ent{Y_1|Y_2^n}.
 \end{equation}
 We can now apply~\cite[Lem.~4.4.1]{Cover_Information1} to both sides to get $\entrate{\Yvec}\ge\ent{Y_2|X_1}$ and $\entrate{\Yvec}\ge\ent{Y_1|X_2}$. We use this in the derivative to get
 \begin{align}
  &\frac{\diff}{\diff\beta}\delta_\beta(\Xvec,\Wvec) \notag\\
  &\ge \mutinf{X_1;X_2}-\ent{Y}+\ent{Y_1|X_2} - \ent{Y_2|Y_1} + \ent{Y_2|X_1}\\
  &= \ent{X|Y} - \ent{X_1|Y_1,X_2} - \ent{Y_2|Y_1} + \ent{Y_2|X_1}\\
  &= \mutinf{X_1;X_2|Y_1} - \mutinf{X_1;Y_2|Y_1}\ge 0
 \end{align}
 by data processing.

\subsection{Proof of Lemma~\ref{lem:costbeta_properties}}\label{proof:costbeta_properties}
The first property follows by recognizing that
\ifCLASSOPTIONdraftcls
	\begin{equation}\label{eq:deltaIsPositive}
	 	\costbeta = (1-\beta) \costl + \beta(\costp-\costl)
	\end{equation}
\else
	\begin{multline}\label{eq:deltaIsPositive}
	\costbeta \\= (1-\beta) \costl + \beta(\costp-\costl)
	\end{multline}
\fi
and that $\costp\ge\costl$.

The second property follows immediately from the definition of $\delta_\beta(\Xvec,\Wvec)$ and $\costp$.

For the third property, note that
\begin{align*}
 \cost_1(\Xvec,\Wvec)
 &=\costp-\costl\\
 &=\mutinf{X_1;X_2}-\ent{Y}+\ent{Y_2|X_1}\\
 &=\mutinf{X_1;X_2}-\mutinf{X_1,Y_2}=\mutinf{X_1;X_2|Y_2}.
\end{align*}

The fourth property is obtained by observing that, if $\beta\le 0.5$
\begin{align*}
 &\delta_\beta(\Xvec,\Wvec)-\beta\mutinf{X_1;X_2} \\
 &= (1-\beta)\ent{Y_2|Y_1}-(1-2\beta)\entrate{\Yvec}-\beta\ent{Y}\\
 &\le (1-\beta)\ent{Y_2|Y_1}-(1-2\beta)\ent{Y_2|X_1}-\beta\ent{Y}\\
 &= (1-2\beta)\ent{Y_2|Y_1}-(1-2\beta)\ent{Y_2|X_1}-\beta\mutinf{Y_1;Y_2}\\
 &= (1-2\beta)\costl -\beta\mutinf{Y_1;Y_2}.
\end{align*}
The inequality is reversed for  $\beta\ge 0.5$.

For the fifth property, we repeat the last steps with 
\begin{equation}
 -(1-2\beta)\entrate{\Yvec}\le-(1-2\beta)\ent{Y_2|Y_1}
\end{equation}
noticing that $(1-2\beta)\le 0$ if $\beta\ge 0.5$. Again, the inequality is reversed for $\beta\le 0.5$.

If $\Xvec$ is reversible, then the PMFs do not change if the order of the indices is reversed. As a consequence, we have $\mutinf{X_1;X_2|Y_2}=\mutinf{X_2;X_1|Y_1}=\cost_1(\Xvec,\Wvec)$. But $\cost_L(\Xvec;\Wvec)=\mutinf{Y_2;X_1|Y_1}\le\cost_1(\Xvec,\Wvec)$ by data processing. Thus, the sixth property follows by noting that, with~\eqref{eq:deltaIsPositive}, $\costbeta=(1-\beta) \costl + \beta\cost_1(\Xvec,\Wvec)$.

\section*{Acknowledgments}
The authors thank Matthias Rungger and Majid Zamani, both from Hybrid Control Systems Group, Technical University of Munich, for discussions suggesting the connection between lumpability and bisimulation.
The work of Rana Ali Amjad was supported by the German Ministry of Education and Research in the framework of an Alexander von Humboldt Professorship.
The work of Bernhard C. Geiger was funded by the Erwin Schr\"odinger Fellowship J 3765 of the Austrian Science Fund.

\bibliographystyle{IEEEtran}
\bibliography{IEEEabrv,../references}

\begin{thebibliography}{10}
\providecommand{\url}[1]{#1}
\csname url@samestyle\endcsname
\providecommand{\newblock}{\relax}
\providecommand{\bibinfo}[2]{#2}
\providecommand{\BIBentrySTDinterwordspacing}{\spaceskip=0pt\relax}
\providecommand{\BIBentryALTinterwordstretchfactor}{4}
\providecommand{\BIBentryALTinterwordspacing}{\spaceskip=\fontdimen2\font plus
\BIBentryALTinterwordstretchfactor\fontdimen3\font minus
  \fontdimen4\font\relax}
\providecommand{\BIBforeignlanguage}[2]{{%
\expandafter\ifx\csname l@#1\endcsname\relax
\typeout{** WARNING: IEEEtran.bst: No hyphenation pattern has been}%
\typeout{** loaded for the language `#1'. Using the pattern for}%
\typeout{** the default language instead.}%
\else
\language=\csname l@#1\endcsname
\fi
#2}}
\providecommand{\BIBdecl}{\relax}
\BIBdecl

\bibitem{Katsoulakis_CoarseGraining}
M.~A. Katsoulakis and J.~Trashorras, ``{Information loss in coarse-graining of
  stochastic particle dynamics},'' \emph{J. Stat. Phys.}, vol. 122, no.~1, pp.
  115--135, 2006.

\bibitem{Abate_Bisimulation}
A.~Abate, ``Approximation metrics based on probabilistic bisimulations for
  general state-space {Markov} processes: {A} survey,'' \emph{Electronic Notes
  in Theoretical Computer Science}, vol. 297, pp. 3 -- 25, 2013, {P}roc.
  Workshop on Hybrid Autonomous Systems.

\bibitem{Alush_PairwiseClustering}
A.~Alush, A.~Friedman, and J.~Goldberger, ``Pairwise clustering based on the
  mutual-information criterion,'' \emph{Neurocomputing}, vol. 182, pp.
  284--293, 2016.

\bibitem{Tishby_MarkovRelaxation}
\BIBentryALTinterwordspacing
N.~Tishby and N.~Slonim, ``Data clustering by {Markovian} relaxation and the
  information bottleneck method,'' in \emph{Advances in Neural Information
  Processing Systems (NIPS)}, Denver, CO, Nov. 2000. [Online]. Available:
  \url{http://citeseer.ist.psu.edu/viewdoc/summary?doi=10.1.1.24.3488}
\BIBentrySTDinterwordspacing

\bibitem{Meyn_MarkovAggregation}
K.~Deng, P.~G. Mehta, and S.~P. Meyn, ``Optimal {Kullback}-{Leibler}
  aggregation via spectral theory of {Markov} chains,'' \emph{{IEEE} Trans.
  Autom. Control}, vol.~56, no.~12, pp. 2793--2808, Dec. 2011.

\bibitem{Xu_Reduction}
Y.~Xu, S.~M. Salapaka, and C.~L. Beck, ``Aggregation of graph models and
  {Markov} chains by deterministic annealing,'' \emph{{IEEE} Trans. Autom.
  Control}, vol.~59, no.~10, pp. 2807--2812, Oct. 2014.

\bibitem{GeigerEtAl_OptimalMarkovAggregation}
B.~C. Geiger, T.~Petrov, G.~Kubin, and H.~Koeppl, ``Optimal
  {Kullback}-{Leibler} aggregation via information bottleneck,'' \emph{{IEEE}
  Trans. Autom. Control}, vol.~60, no.~4, pp. 1010--1022, Apr. 2015,
  open-access: {\tt arXiv:1304.6603 [cs.SY]}.

\bibitem{Vidyasagar_MarkovAggregation}
M.~Vidyasagar, ``Reduced-order modeling of {Markov} and hidden {Markov}
  processes via aggregation,'' in \emph{Proc. IEEE Conf. on Decision and
  Control (CDC)}, Atlanta, GA, Dec. 2010, pp. 1810--1815.

\bibitem{Kemeny_FMC}
J.~G. Kemeny and J.~L. Snell, \emph{Finite Markov Chains}, 2nd~ed.\hskip 1em
  plus 0.5em minus 0.4em\relax Springer, 1976.

\bibitem{Gray_Entropy}
R.~M. Gray, \emph{Entropy and Information Theory}.\hskip 1em plus 0.5em minus
  0.4em\relax New York, NY: Springer, 1990.

\bibitem{Rached_KLDR}
Z.~Rached, F.~Alajaji, and L.~L. Campbell, ``The {Kullback-Leibler} divergence
  rate between {Markov} sources,'' \emph{{IEEE} Trans. Inf. Theory}, vol.~50,
  no.~5, pp. 917--921, May 2004.

\bibitem{GeigerTemmel_kLump}
B.~C. Geiger and C.~Temmel, ``Lumpings of {Markov} chains, entropy rate
  preservation, and higher-order lumpability,'' \emph{J. Appl. Probab.},
  vol.~51, no.~4, pp. 1114--1132, Dec. 2014, extended version available: {\tt
  arXiv:1212.4375 [cs.IT]}.

\bibitem{Blackwell_HMMRate}
D.~Blackwell, ``{The entropy of functions of finite-state \{M\}arkov chains},''
  in \emph{Trans. first \{P\}rague Conf. Inf. theory, \{S\}tatistical Decis.
  Funct. random Process. held \{L\}iblice near \{P\}rague from \{N\}ovember 28
  to 30, 1956}.\hskip 1em plus 0.5em minus 0.4em\relax Prague: Publishing House
  of the Czechoslovak Academy of Sciences, 1957, pp. 13--20.

\bibitem{Ordentlich_LowerBound}
O.~Ordentlich, ``Novel lower bounds on the entropy rate of binary hidden
  {Markov} processes,'' in \emph{Proc. IEEE Int. Sym. on Information Theory
  (ISIT)}, Jul. 2016, pp. 690--694.

\bibitem{Desharnais_Bisimulation}
J.~Desharnais, A.~Edalat, and P.~Panangaden, ``Bisimulation for labelled
  {Markov} processes,'' \emph{Information and Computation}, vol. 179, no.~2,
  pp. 163 -- 193, 2002.

\bibitem{Bian_Bisimulation}
G.~Bian and A.~Abate, ``On the relationship between bisimulation and trace
  equivalence in an approximate probabilistic context,'' in \emph{Proc. Int.
  Conf. on Foundations of Software Science and Computation Structure
  (FOSSACS)}, J.~Esparza and A.~S. Murawski, Eds.\hskip 1em plus 0.5em minus
  0.4em\relax Uppsala: Springer Berlin Heidelberg, Apr. 2017, pp. 321--337.

\bibitem{GeigerAmjad_HardClusters}
B.~C. Geiger and R.~A. Amjad, ``Mutual information-based clustering: Hard or
  soft?'' in \emph{Proc. Int. ITG Conf. on Systems, Communications and Coding
  (SCC)}, Hamburg, Feb. 2017, pp. 1--6, open-access: {\tt arXiv:1608.04872
  [cs.IT]}.

\bibitem{Tishby_InformationBottleneck}
N.~Tishby, F.~C. Pereira, and W.~Bialek, ``The information bottleneck method,''
  in \emph{Proc. Allerton Conf. on Communication, Control, and Computing},
  Monticello, IL, Sep. 1999, pp. 368--377.

\bibitem{Slonim_PhD}
N.~Slonim, ``The information bottleneck: Theory and applications,'' Ph.D.
  dissertation, Hebrew University of Jerusalem, 2002.

\bibitem{Slonim_Agglomerative}
N.~Slonim and N.~Tishby, ``Agglomerative information bottleneck,'' in
  \emph{Advances in Neural Information Processing Systems (NIPS)}, Denver, CO,
  Nov. 1999, pp. 617--623.

\bibitem{blakegraduatedopt}
A.~Blake and A.~Zisserman, \emph{Visual Reconstruction}.\hskip 1em plus 0.5em
  minus 0.4em\relax Cambridge, MA, USA: MIT Press, 1987.

\bibitem{Cover_Information1}
T.~M. Cover and J.~A. Thomas, \emph{Elements of Information Theory},
  1st~ed.\hskip 1em plus 0.5em minus 0.4em\relax Wiley Interscience, 1991.

\bibitem{GeigerWu_HigherOrder}
B.~C. Geiger and Y.~Wu, ``Higher-order optimal {K}ullback-{L}eibler aggregation
  of {M}arkov chains,'' in \emph{Proc. Int. ITG Conf. on Systems,
  Communications and Coding (SCC)}, Hamburg, Feb. 2017, pp. 1--6, open-access:
  {\tt arXiv:1608.04637 [cs.IT]}.

\end{thebibliography}

\end{document}